\documentclass[%
 reprint,
 superscriptaddress,
 amsmath,amssymb,
 aps, pra,
 floatfix
]{revtex4-2}

\usepackage[utf8]{inputenc}
\usepackage[T1]{fontenc}
\usepackage[english]{babel}

\usepackage{graphicx}

\makeatletter
\renewcommand{\fnum@figure}{FIG.\ \thefigure} 
\makeatother

\usepackage{microtype}
\usepackage{anyfontsize}
\usepackage[varvw]{newtx}
\usepackage{soul}

\usepackage{mathtools}
\usepackage[separate-uncertainty=true, text-series-to-math=true, propagate-math-font=true]{siunitx}
\usepackage{braket}
\usepackage{bm}

\usepackage{booktabs}

\usepackage[dvipsnames]{xcolor}
\usepackage[hidelinks,colorlinks=true,allcolors=Blue]{hyperref}
\usepackage[capitalize]{cleveref}

\newcommand{\commentblock}[1]{}

\def\i{\mathrm i}

\begin{document}

\hyphenpenalty=9999

\preprint{APS/123-QED}

\title{50-km fiber interferometer for testing gravitational signatures in quantum interference}

\author{Haocun~Yu} \email{haocunyu@utk.edu}
\altaffiliation{Current affiliation: University of Tennessee, Knoxville, Department of Physics and Astronomy, 408 Circle Dr, Knoxville, TN 37996, USA}
\affiliation{University of Vienna, Faculty of Physics, \\Vienna Center for Quantum Science and Technology (VCQ), \\Boltzmanngasse 5, 1090 Vienna, Austria}
\affiliation{University of Vienna, Research Network Quantum Aspects of Spacetime (TURIS), \\Boltzmanngasse 5, 1090 Vienna, Austria}

\author{Dorotea~Macri}
\affiliation{Massachusetts Institute of Technology, Cambridge, Massachusetts 02139, USA}

\author{Thomas~Morling}
\affiliation{University of Vienna, Faculty of Physics, \\Vienna Center for Quantum Science and Technology (VCQ), \\Boltzmanngasse 5, 1090 Vienna, Austria}
\affiliation{University of Vienna, Faculty of Physics and Vienna Doctoral School in Physics (VDSP), \\Boltzmanngasse 5, 1090 Vienna, Austria}
\affiliation{University of Vienna, Research Network Quantum Aspects of Spacetime (TURIS), \\Boltzmanngasse 5, 1090 Vienna, Austria}

\author{Eleonora~Polini}
\altaffiliation{Current affiliation: Université Côte d’Azur, Observatoire de la Côte d’Azur, Artemis, CNRS, 06304 Nice, France}
\affiliation{Massachusetts Institute of Technology, Cambridge, Massachusetts 02139, USA}

\author{Thomas~B.~Mieling}
\affiliation{University of Vienna, Faculty of Physics, \\Vienna Center for Quantum Science and Technology (VCQ), \\Boltzmanngasse 5, 1090 Vienna, Austria}
\affiliation{University of Vienna, Research Network Quantum Aspects of Spacetime (TURIS), \\Boltzmanngasse 5, 1090 Vienna, Austria}

\author{Peter~Barrow}
\affiliation{University of Vienna, Faculty of Physics, \\Vienna Center for Quantum Science and Technology (VCQ), \\Boltzmanngasse 5, 1090 Vienna, Austria}
\affiliation{University of Vienna, Research Network Quantum Aspects of Spacetime (TURIS), \\Boltzmanngasse 5, 1090 Vienna, Austria}

\author{Begüm~Kabagöz}
\affiliation{Massachusetts Institute of Technology, Cambridge, Massachusetts 02139, USA}

\author{Xinghui~Yin}
\affiliation{Massachusetts Institute of Technology, Cambridge, Massachusetts 02139, USA}

\author{Piotr~T.~Chruściel}
\affiliation{Center for Theoretical Physics of the Polish Academy of Sciences, \\al.\ Lotnik\'ow 32/46, PL 02-668 Warszawa, Poland}

\author{Christopher~Hilweg}
\affiliation{University of Vienna, Faculty of Physics, \\Vienna Center for Quantum Science and Technology (VCQ), \\Boltzmanngasse 5, 1090 Vienna, Austria}
\affiliation{University of Vienna, Research Network Quantum Aspects of Spacetime (TURIS), \\Boltzmanngasse 5, 1090 Vienna, Austria}

\author{Eric~Oelker}
\affiliation{Massachusetts Institute of Technology, Cambridge, Massachusetts 02139, USA}

\author{Nergis~Mavalvala}
\affiliation{Massachusetts Institute of Technology, Cambridge, Massachusetts 02139, USA}

\author{Philip~Walther}
\email{philip.walther@univie.ac.at}
\affiliation{University of Vienna, Faculty of Physics, \\Vienna Center for Quantum Science and Technology (VCQ), \\Boltzmanngasse 5, 1090 Vienna, Austria}
\affiliation{University of Vienna, Research Network Quantum Aspects of Spacetime (TURIS), \\Boltzmanngasse 5, 1090 Vienna, Austria}
\affiliation{Institute for Quantum Optics and Quantum Information (IQOQI) Vienna, Austrian Academy of Sciences, Boltzmanngasse 3, 1090 Vienna, Austria}


\begin{abstract}

Quantum mechanics and general relativity are the foundational pillars of modern physics, yet experimental tests that combine the two frameworks remain rare. 
Measuring optical phase shifts of massless photons in a gravitational potential provides a unique quantum platform to probe gravity beyond Newtonian descriptions, but laboratory-based interferometers have not yet reached the sensitivity needed to access this regime. 
Here, we report the realization of a 50-km table-top Mach-Zehnder fiber interferometer operating at the single-photon level, achieving a phase sensitivity of $\SI{4.42e-6}{\radian}$ root-mean-square (RMS) within the frequency range of \SIrange{0.01}{5}{\hertz}.
We demonstrate that this sensitivity is sufficient to resolve a phase-shift signal of \SI{6.18(44)e-5}{\radian} RMS at \SI{0.1}{\hertz}, associated with a modulated gravity-induced signal.
Our results establish a milestone for quantum sensing with large-scale optical interferometry, demonstrating the capability to detect gravitational redshifts in a local laboratory, thereby paving the way for testing quantum phenomena within general relativistic frameworks.

\end{abstract}

\maketitle


\begin{figure*}[t]
    \centering
    \includegraphics[width=\textwidth]{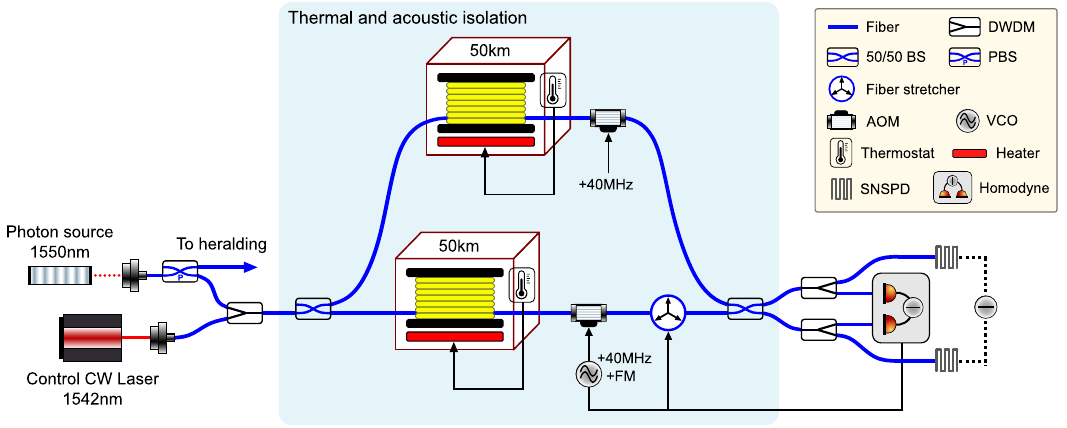}
    \caption{
    Layout of the Mach-Zehnder fiber interferometer, with both interferometer arms maintained at the same height.
    The interferometer is composed of two 50/50 fiber beam splitters (BS) and two 50-km low-loss fiber spools (yellow), each maintained under active temperature control. Single photons centered at \SI{1550}{\nano\meter}, generated from a Type-0 photon source, are used to probe the interferometer phase sensitivity. 
    A weak continuous-wave (CW) laser field at \SI{1542}{\nano\meter} is used to lock the interferometer phase. This classical field co-propagates with the single photons through the interferometer before being separated by dense wavelength division multiplexers (DWDM) and directed to a homodyne detector. 
    To stabilize both fast and slow phase fluctuations, the error signal from the homodyne detector is sent to an acousto-optic modulator (AOM) driven by a voltage-controlled oscillator (VCO) and to a fiber stretcher, respectively.
    The interferometer phase is extracted from the heralded photon number counts detected by superconducting nanowire single-photon detectors (SNSPDs). 
    The core interferometer elements are placed in a thermally and acoustically isolated enclosure (blue shaded region).
    }
    \label{fig:exp_setup detailed}
\end{figure*}

\begin{figure*}[t]
    \centering\includegraphics[width=0.9\textwidth]{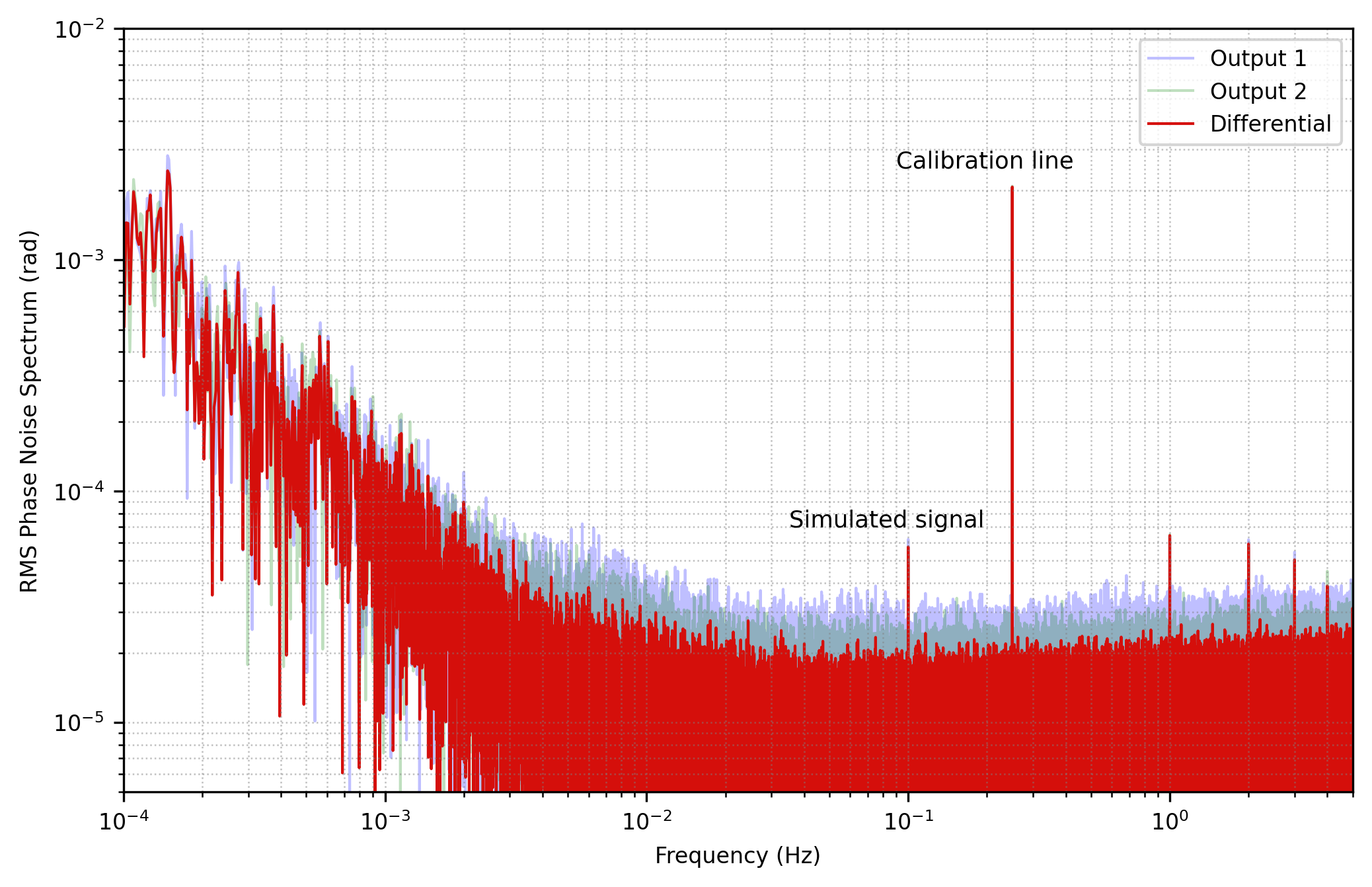}
    \caption{
    Phase noise spectra of the single-photon interferometer from a 160-hour measurement, spanning the frequency band from \SI{e-4}{\hertz} to \SI{5}{\hertz}. The dim blue and green traces show the phase noise sensitivity calibrated from heralded single-photon counts at each interferometer output.
    The red trace shows the half-difference between the phase noises measured at the two interferometer outputs, rejecting common-mode noise of the interferometer. The peak at \SI{0.25}{\hertz} corresponds to the injected dither tone for calibration; 
    The peak observed at \SI{0.1}{\hertz} corresponds to the simulated gravitationally-induced phase shift signal, with an expected root-mean-square (RMS) amplitude of \SI{6.48e-5}{\radian}, in agreement with the measured value of \SI{6.18(44)e-5}{\radian}.
    The signal is clearly resolved above the noise floor, which is primarily limited by photon shot noise.
    Additional peaks at \SI{1}{\hertz} and its harmonics arise from cryostat compressor noise.}
    \label{fig:photon_spec}
\end{figure*}

\begin{figure*}[t]
    \centering
    \includegraphics[width=\textwidth]{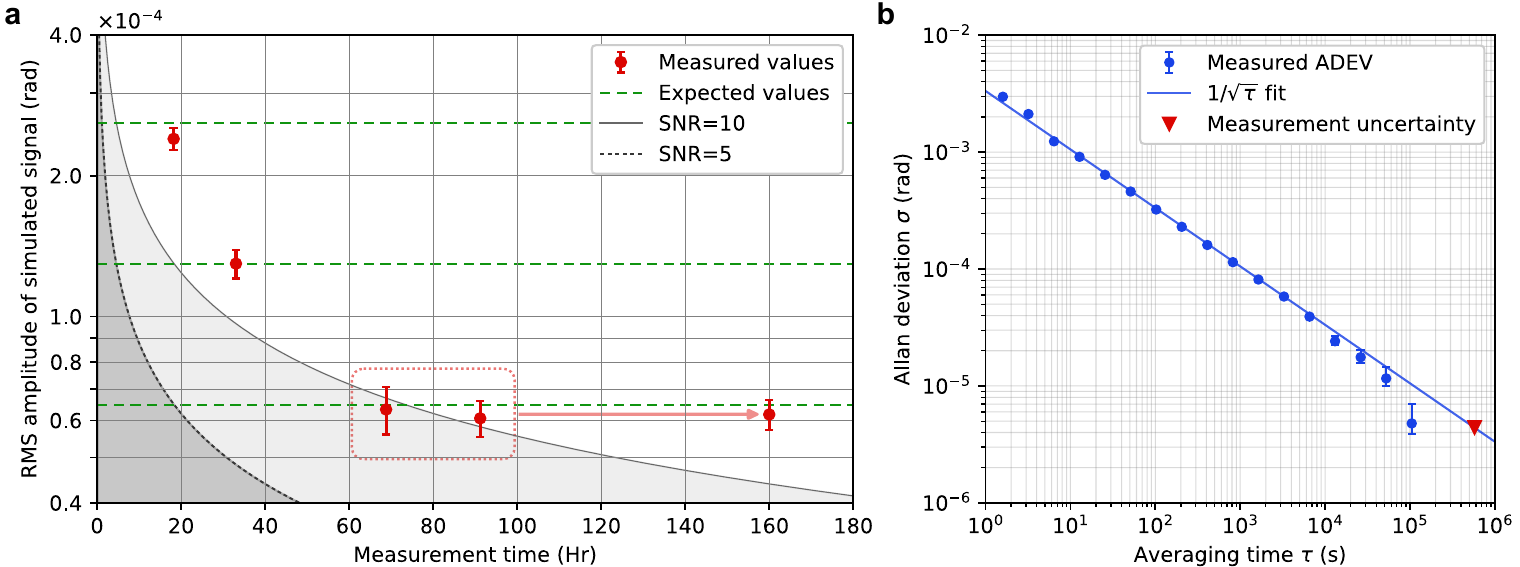}
    \caption{
    (a) Measurements of injected signal amplitudes at \SI{0.1}{\hertz}. 
    We use three different dithers with calibrated RMS amplitudes of \num{2.59e-04}, \num{1.30e-04}, and \SI{6.48e-05}{\radian} (green dashed lines).
    The measurement durations are 18.2, 33.1, and 160 hours, respectively.
    The measured signal values are \num{2.40(13)e-4}, \num{1.30(09)e-4}, \SI{6.18(44)e-5}{\radian} (red dots with error bars).
    The data for the smallest dither was combined from two separate runs (68.8~h and 91.2~h), yielding independent estimates of \num{6.33(73)e-5} and \SI{6.06(54)e-5}{rad} (red dotted box). This demonstrates the repeatability of the measurement and the stability of the apparatus.
    The thresholds for signal-to-noise ratios (SNR) of 5 and 10 are shown in the plot to indicate the required measurement times for different signal levels.
    (b) Allan deviation (ADEV) of the in-phase component obtained from lock-in analysis of the simulated signal, showing the phase stability $\sigma$ versus the averaging time $\tau$. 
    The blue dots with error bars show overlapping ADEV values calculated from the 160-hour measurement.
    The data follows a $1/\sqrt{\tau}$ trend (blue line) and is consistent with a white noise process, indicating that the measured signal is stationary. The red triangle shows an estimate statistical uncertainty of the measurement from the demodulated time series.  It is consistent with an extrapolation of the ADEV trend, as expected for a white noise process.}
    \label{fig:uncertainty}
\end{figure*}

\textit{Introduction.---}
Quantum mechanics revolutionized our understanding of the microscopic world, while Einstein's theory of general relativity provides the most accurate description of gravity at macroscopic, even cosmological, scales. 
However, a unified framework connecting these two pillars of modern physics remains elusive, hindered by the lack of experimental insight.
Although both theories have been tested with high precision~\cite{Hanneke2008,Will2014,ligo_2016,Delva_2018, Takamoto_2020, Bothwell_2022, zheng_2023}, no experiment has yet accessed a regime where their concepts must be applied simultaneously.
In recent decades, numerous proposals and experiments have sought to test gravitational effects on quantum systems. These efforts fall broadly into two categories: investigations of the emergence of entanglement arising solely from gravitational coupling between particles~\cite{Bose2017,Marletto2017,Hall2018,Kent2021}, and studies of how quantum states are influenced by classical gravitational fields~\cite{COW_1975_PRL, Kasevich1991,Kasevich1992,Snadden1998,Altin2013,Xu2019,Panda2024,Overstreet2022}. 
Since the first observation of gravitationally-induced phase shifts in neutrons~\cite{COW_1975_PRL}, the latter approach has seen significant progress with matter wave interferometers~\cite{Kasevich1991,Kasevich1992,Snadden1998,Altin2013,Xu2019,Panda2024,Overstreet2022}.

However, all experiments conducted to date remain consistent with Newtonian gravity and, therefore, do not necessitate a general relativistic explanation.
Photons, as massless quantum particles, provide an ideal platform to probe the influence of gravity on quantum systems in regimes that elude a Newtonian description, thereby necessitating the simultaneous inclusion of quantum mechanics and general relativity~\cite{Zych_2012, MMHW_2025}.

Since general relativistic effects are extremely small in Earth’s weak gravitational field, experiments involving both general relativity and quantum mechanics are necessarily at the cutting edge of precision measurement.
A promising approach to exploring this interface is to employ long-baseline photonic interferometers, in which the extended photon propagation time at different heights allows weak gravitational effects to build up to measurable levels.
Several satellite-based proposals and initial proof-of-principle experiments have been put forward to realize this vision~\cite{Rideout2012,vallone_2016,Ping_2019,RMP_Micius_2022,2022PhRvD.105j5016B,Mieling2022,Mohageg2022,Wu_2024_PRL}. 
In practice, space-based experiments are technically demanding~---~fragile, costly, and often constrained by local weather and atmospheric conditions.

Fiber-based interferometers, on the other hand, provide a practical means to obtaining long photon propagation times~---~and therefore measurable optical phase shifts~---~within compact and controllable laboratory environments ~\cite{Stodolsky1979,Tanaka1983,Zych_2012,Hilweg_2017,Chen2019,Barzel2024}.
However, despite significant advances in optical platforms for quantum information processing~\cite{RevModPhys.74.145,RevModPhys.79.135,OBrien_2009} and quantum networks~\cite{Kimble_2008, Wehner_2018, RevModPhys.92.025002}, laboratory-scale optical interferometry experiments have not yet achieved the sensitivity required to probe gravitational signatures. 
Recent experimental endeavors~\cite{Silvestri_2024} toward this goal remain limited by pervasive environmental noise sources~\cite{Hilweg_2022}.
We aim to access the regime of measurable gravitational effects on photonic states in the context of general relativity, by performing precision measurements using a large-scale, tabletop fiber interferometer whose arms are positioned at different heights to experience distinct gravitational potentials~\cite{ERC}.

In a Mach-Zehnder interferometer consisting of two fiber spools of equal length $l$ separated vertically by a height $h$, the expected gravitational phase shift for a single photon of vacuum wavelength $\lambda$ is
\begin{align}
    \label{eq:gravitational phase shift}
    \begin{split}
        \Delta \phi_g
            &= 2 \pi n g h l / (\lambda c^2)
            \\&
            \approx \SI{e-6}{\radian} \times \left(\frac{h}{\si{\meter}}\right) \left(\frac{l}{\si{\kilo\meter}}\right) \left(\frac{\lambda}{\si{\micro\meter}}\right)^{-1}
            \,,
    \end{split}
\end{align}
where $n \approx \num{1.46}$ is the effective refractive index of the silica optical fiber, $g$ is the local gravitational acceleration and $c$ is the speed of light in vacuum \cite{Mieling_2025,MMHW_2025}. 
In this work, we report our first realization of a laboratory-based fiber interferometer with two arms of length $l = \SI{50}{\kilo\meter}$ located at the same height ($h = 0$), utilizing single-photon interference to achieve a phase sensitivity of $\SI{4.42e-6}{\radian}$ root-mean-square (RMS) within the frequency range of \SIrange{0.01}{5}{\hertz}.
We demonstrate that the achieved sensitivity is sufficient to clearly resolve a simulated signal with a known magnitude of $\SI{6.48e-5}{\radian\,RMS}$, comparable to the gravitational phase shift accessible with a tabletop apparatus.
This surpasses the sensitivity of previous single-photon fiber interferometers by two orders of magnitude~\cite{Silvestri_2024} and paves the way for testing quantum phenomena in the context of general relativity in a local laboratory.

\textit{Experimental Setup.---}
A sketch of our experimental setup is shown in \cref{fig:exp_setup detailed}. 
The apparatus consists of a bright single-photon source, a Mach-Zehnder interferometer with \SI{50}{\kilo\meter}-long arms built entirely from fiber-based components, superconducting nanowire single-photon detectors (SNSPD), and an auxiliary system employing continuous-wave (CW) control laser light with balanced homodyne detection to stabilize the interferometer.
Broadband single photons centered at \SI{1550.12}{\nano\meter} are generated via a Type-0 spontaneous parametric down-conversion process in a periodically poled lithium niobate (PPLN) nonlinear crystal~\cite{Neumann_2022, Ortega:23}. A polarizing beam splitter (PBS) separates the generated photon pairs, sending one photon through the interferometer while the other serves as a herald and is sent directly to the detector.
(For details of the photon source, see Supplementary Materials~\cite{sm}.)
To strike a balance between a feasible coherence length and an adequate count rate for a reasonable integration time, the photons are filtered to a bandwidth of \SI{100}{\giga\hertz} using dense wavelength division multiplexers (DWDM), resulting in single photon counts exceeding \SI{40}{\mega\hertz}.

The interferometer arms are composed of two compact fiber spools, each housed in a thermally insulated cylindrical enclosure with active temperature feedback control, providing stability at the \SI{0.1}{\milli\kelvin} level to minimize phase noise from thermal fluctuations.
We use a separate pre-stabilized laser with a 1-Hz linewidth at a wavelength of \SI{1542.14}{\nano\meter} as a control field to actively stabilize the interferometer phase.
The laser power is minimized to prevent spontaneous Raman-scattered photons from contaminating the single-photon channel and saturating the SNSPDs~\cite{mlejnek_2017,KAWAHARA_2011}.
This control light field is combined with single photons using a DWDM, co-propagates through the interferometer, and is then separated and directed to a balanced homodyne detector upon exiting.
Error signals from the homodyne detection are fed back to an acousto-optic modulator (AOM) and a piezo-electrically actuated fiber stretcher for phase correction. This control field is also used to characterize the noise performance of the interferometer in the classical regime.
To passively mitigate environmental noise, the majority of the interferometer components are enclosed within a thermally and acoustically insulated housing (blue shaded region in \cref{fig:exp_setup detailed}). 

After interfering at the second beam splitter, the single photons leaving the interferometer are measured with single pixel SNSPDs, while the heralding photons are monitored with multipixel SNSPDs.
In this current setup, we inject weak sinusoidal modulations into the control signal to introduce $S_g$, a signal associated with the gravitationally-induced phase shift $\Delta\phi_g$.
The total interferometer phase $\phi_0+S_g$ is then extracted from the heralded photon counts at each output port, given by $N_{1,2} = \tfrac{1}{2} N \big[ 1 \pm V \cos(\phi_0+S_g) \big]$, where $N$ denotes the total photon counts out of the interferometer, $V$ the fringe visibility, and $\phi_0$ the initial locking offset phase of the interferometer~\cite{Zych_2012, MMHW_2025}. 

\textit{Experimental Results.---}
This experiment aims to demonstrate the highest phase sensitivity achieved in a laboratory-based large-scale fiber interferometer, and to showcase its capability to resolve the minute signal $S_g$. 
Our measurements comprise two long data runs of \num{68.8} and \num{91.2} hours. The experimental conditions across the two runs were sufficiently stable and consistent to justify stitching the data into a single 160-hour dataset, referred to as the “main measurement,” shown in Fig.~\ref{fig:photon_spec} and \hyperref[fig:uncertainty]{3(a)}.
To support our analyses we also carry out a series of auxiliary measurements, shown in Fig.~\hyperref[fig:uncertainty]{3(b)}.

\Cref{fig:photon_spec} presents the single-photon phase noise amplitude spectra after 160 hours, revealing the emergence of the simulated signal $S_g$ from the background noise.
The interferometer is locked near $\phi_0=\pi/2$ with photon interference visibility $V=98\%$. Two sinusoidal dithers are injected into the feedback control loop to characterize the noise performance more accurately --- a reference tone with a known RMS amplitude of \SI{2.10e-3}{\radian} at \SI{0.25}{\hertz} used to refine the calibration, and the signal tone $S_g$ at \SI{0.1}{\hertz}.
In the main measurement, the RMS amplitude of $S_g$ was set to \SI{6.48e-5}{\radian}.

The time series data of heralded photon counts from the two interferometer outputs are acquired at a sampling rate of \SI{10}{\hertz} using a time tagger. 
The measured photon counts are calibrated to RMS phase noise in radians based on the fringe visibility of the interferometer determined from a full phase scan.
To reject common-mode phase noise, we calculate the half-difference between the phase signals extracted from the two interferometer output ports, as shown by the red trace.
The spectra are computed with the finest available frequency resolution to maximize the contrast between the signal and background noise.

The achieved level of sensitivity allows the injected signal to emerge distinctly above the background phase noise. 
The interferometer noise above \SI{0.01}{\hertz} is limited by photon shot noise, reaching a level of \SI{4.42e-3}{\radian\per\sqrt{\hertz}}.
This corresponds to a fractional displacement sensitivity, or a travel time dilation for \SI{1550}{\nano\meter} photons, of \SI{1.52e-14}{\text{/}\sqrt{\hertz}}.
At lower frequencies, classical noise dominates due to residual thermal fluctuations, seismic vibrations, and acoustic disturbances coupling into fiber phase noise.
A full characterization of the interferometer using classical light, including the photon noise spectral density and its comparison to the classical noise measurement can be found in Supplementary Material and Fig.~S2.

A post-processing lock-in analysis is applied to eliminate $\phi_0 $ while extracting the amplitude and uncertainty of $S_g$ more precisely.
The in-phase components of the electrical signal, demodulated at \SI{0.25}{\hertz} and \SI{0.1}{\hertz}, are extracted after low-pass filtering.
During the extended measurement period, the visibility of the interferometer degrades due to polarization and temperature drifts.
To compensate for this, we divide each of the two data series into 10 segments and rescale the calibration factor in each segment by matching the \SI{0.25}{\hertz} dither amplitude to the known level of the calibration signal.
With the calibration applied, we compute the mean of the demodulated time-series and quantify its uncertainty using the standard error of the mean, yielding the measured signal of $S_g$ = \SI{6.18(44)e-5}{\radian}, which encompasses the expected value within its uncertainty. 
\Cref{fig:uncertainty}(a) shows the overlapping Allan deviation (ADEV) plot, which characterizes the stability of the measured phase shift over an averaging period $\tau$.
The ADEV trend (fitted blue line) is consistent with a stationary noise process, indicating that the interferometer’s mean phase shift remains constant throughout the measurement.
By extrapolating this trend to the full measurement duration ($\tau$ = 160 hours), we see that it is consistent with our estimate of the statistical uncertainty (red triangle).
The time-series data plot from the lock-in analysis can be found in Supplementary Material Fig.~S3.

To assess the repeatability and consistency of the results, we performed $S_g$ injections and measurements over a range of amplitudes with varying integration times.
The measurement results for the three different signal levels are summarized and presented in \cref{fig:uncertainty}(b). 
For each injection, we extracted the signal amplitude and uncertainty following the same procedure used for the main measurement.
\cref{fig:uncertainty}(b) also indicates the measurement integration time necessary to reach a sufficiently high signal-to-ratio noise (SNR) for signals of different amplitudes.

\begin{table}[htb]
    \centering
    \caption{Measured loss budget from various sources and the total optical loss of the entire fiber interferometer setup, from the photon source output to the detection stage. Measurements were performed using CW laser light at a wavelength of \SI{1550}{\nano\meter}.}
    \label{tab:table1}
    \vspace{1.5\baselineskip}
    \begin{tabular}{@{}llr@{}}
        \toprule
        Loss source &   & Loss (dB) \\ 
        \midrule
        Fiber spool &  & \num{9.75(01)} \\
        Optical components    
        & AOM          & \num{2.35(02)}  \\ 
        & DWDMs        & \num{1.60(02)}  \\ 
        & BSs          & \num{0.09(01)}  \\ 
        Fiber connections && \num{1.00(02)} \\ 
        SNSPD efficiency && \num{0.20(02)}  \\ 
        \midrule
        \textbf{Total} && \textbf{\num{14.99(10)}}  \\
        \bottomrule
    \end{tabular}
    \vspace{-\baselineskip}
\end{table}

\textit{Conclusions.---}
In summary, we have demonstrated a 50-km fiber interferometer that, for the first time, achieves the sensitivity needed to observe gravitational redshifts with a photonic quantum state under laboratory conditions.
This achievement establishes a milestone for quantum sensing with optical fiber interferometry, highlighting the potential to explore gravitational effects on single and entangled photons within the framework of general relativity.

Currently, the sensitivity performance is primarily limited by the optical loss of the whole fiber system.
A loss budget from various sources of the setup is summarized in \cref{tab:table1}. 
The total loss of each interferometer path is \SI{14.99(10)}{\dB}, corresponding to a transmission efficiency of \SI{3.17(7)}{\percent}.
This substantial loss reduces the photon count at the outputs, imposing a photon shot-noise limitation at frequencies above \SI{0.01}{\hertz}.
To reduce optical loss, the current fibers can be replaced with ultra-low-loss fibers with spliced connections, minimizing insertion losses. 
Regarding classical noise, seismic disturbances can be further suppressed with vibration-minimizing fiber spools, while long-term phase drift arising from polarization and temperature fluctuations can be mitigated through active polarization control and enhanced thermal insulation.

Based on the results presented here and supported by extensive theoretical studies~\cite{Mieling_2020, Mieling_2022, Mieling_2025}, the authors have initiated the design and construction of a next-generation setup incorporating a modulated vertical height difference between two interferometer arms, along with further technical upgrades. 
This differential measurement scheme (see Supplementary Materials for additional details) will be capable of comparing gravitational redshift-induced effects on classical and quantum light, probing potential new effects of gravity on entangled quantum fields, and taking an important step toward experimentally accessing the interface between quantum mechanics and general relativity.

\begin{acknowledgments}

This work is funded by the European Union (ERC, GRAVITES, No.~101071779). Views and opinions expressed are however those of the author(s) only and do not necessarily reflect those of the European Union or the European Research Council Executive Agency.
H.Y. acknowledges funding from the European Union HORIZON TMA MSCA Postdoctoral Fellowships - European Fellowships under grant agreement no.~101064373 (MAGIQUE).
D.M. is supported by the U.S. National Science Foundation's Graduate Research Fellowship Program.
T.M. is supported by the Vienna Doctoral School in Physics (VDSP).
\end{acknowledgments}

\bibliography{references}

\begin{thebibliography}{51}%
\makeatletter
\providecommand \@ifxundefined [1]{%
 \@ifx{#1\undefined}
}%
\providecommand \@ifnum [1]{%
 \ifnum #1\expandafter \@firstoftwo
 \else \expandafter \@secondoftwo
 \fi
}%
\providecommand \@ifx [1]{%
 \ifx #1\expandafter \@firstoftwo
 \else \expandafter \@secondoftwo
 \fi
}%
\providecommand \natexlab [1]{#1}%
\providecommand \enquote  [1]{``#1''}%
\providecommand \bibnamefont  [1]{#1}%
\providecommand \bibfnamefont [1]{#1}%
\providecommand \citenamefont [1]{#1}%
\providecommand \href@noop [0]{\@secondoftwo}%
\providecommand \href [0]{\begingroup \@sanitize@url \@href}%
\providecommand \@href[1]{\@@startlink{#1}\@@href}%
\providecommand \@@href[1]{\endgroup#1\@@endlink}%
\providecommand \@sanitize@url [0]{\catcode `\\12\catcode `\$12\catcode `\&12\catcode `\#12\catcode `\^12\catcode `\_12\catcode `\%12\relax}%
\providecommand \@@startlink[1]{}%
\providecommand \@@endlink[0]{}%
\providecommand \url  [0]{\begingroup\@sanitize@url \@url }%
\providecommand \@url [1]{\endgroup\@href {#1}{\urlprefix }}%
\providecommand \urlprefix  [0]{URL }%
\providecommand \Eprint [0]{\href }%
\providecommand \doibase [0]{https://doi.org/}%
\providecommand \selectlanguage [0]{\@gobble}%
\providecommand \bibinfo  [0]{\@secondoftwo}%
\providecommand \bibfield  [0]{\@secondoftwo}%
\providecommand \translation [1]{[#1]}%
\providecommand \BibitemOpen [0]{}%
\providecommand \bibitemStop [0]{}%
\providecommand \bibitemNoStop [0]{.\EOS\space}%
\providecommand \EOS [0]{\spacefactor3000\relax}%
\providecommand \BibitemShut  [1]{\csname bibitem#1\endcsname}%
\let\auto@bib@innerbib\@empty
\bibitem [{\citenamefont {Hanneke}\ \emph {et~al.}(2008)\citenamefont {Hanneke}, \citenamefont {Fogwell},\ and\ \citenamefont {Gabrielse}}]{Hanneke2008}%
  \BibitemOpen
  \bibfield  {author} {\bibinfo {author} {\bibfnamefont {D.}~\bibnamefont {Hanneke}}, \bibinfo {author} {\bibfnamefont {S.}~\bibnamefont {Fogwell}},\ and\ \bibinfo {author} {\bibfnamefont {G.}~\bibnamefont {Gabrielse}},\ }\bibfield  {title} {\bibinfo {title} {New measurement of the electron magnetic moment and the fine structure constant},\ }\href {https://doi.org/10.1103/PhysRevLett.100.120801} {\bibfield  {journal} {\bibinfo  {journal} {Phys. Rev. Lett.}\ }\textbf {\bibinfo {volume} {100}},\ \bibinfo {pages} {120801} (\bibinfo {year} {2008})}\BibitemShut {NoStop}%
\bibitem [{\citenamefont {Will}(2014)}]{Will2014}%
  \BibitemOpen
  \bibfield  {author} {\bibinfo {author} {\bibfnamefont {C.~M.}\ \bibnamefont {Will}},\ }\bibfield  {title} {\bibinfo {title} {The confrontation between general relativity and experiment},\ }\href {https://doi.org/10.12942/lrr-2014-4} {\bibfield  {journal} {\bibinfo  {journal} {Living Reviews in Relativity}\ }\textbf {\bibinfo {volume} {17}},\ \bibinfo {pages} {4} (\bibinfo {year} {2014})}\BibitemShut {NoStop}%
\bibitem [{\citenamefont {{LIGO Scientific Collaboration and Virgo Collaboration}}(2016)}]{ligo_2016}%
  \BibitemOpen
  \bibfield  {author} {\bibinfo {author} {\bibnamefont {{LIGO Scientific Collaboration and Virgo Collaboration}}},\ }\bibfield  {title} {\bibinfo {title} {{Observation of Gravitational Waves from a Binary Black Hole Merger}},\ }\href {https://doi.org/10.1103/PhysRevLett.116.061102} {\bibfield  {journal} {\bibinfo  {journal} {Physical Review Letters}\ }\textbf {\bibinfo {volume} {116}},\ \bibinfo {eid} {061102} (\bibinfo {year} {2016})}\BibitemShut {NoStop}%
\bibitem [{\citenamefont {{Delva}}\ \emph {et~al.}(2018)\citenamefont {{Delva}}, \citenamefont {{Puchades}}, \citenamefont {{Sch{\"o}nemann}}, \citenamefont {{Dilssner}}, \citenamefont {{Courde}}, \citenamefont {{Bertone}}, \citenamefont {{Gonzalez}}, \citenamefont {{Hees}}, \citenamefont {{Le Poncin-Lafitte}}, \citenamefont {{Meynadier}}, \citenamefont {{Prieto-Cerdeira}}, \citenamefont {{Sohet}}, \citenamefont {{Ventura-Traveset}},\ and\ \citenamefont {{Wolf}}}]{Delva_2018}%
  \BibitemOpen
  \bibfield  {author} {\bibinfo {author} {\bibfnamefont {P.}~\bibnamefont {{Delva}}}, \bibinfo {author} {\bibfnamefont {N.}~\bibnamefont {{Puchades}}}, \bibinfo {author} {\bibfnamefont {E.}~\bibnamefont {{Sch{\"o}nemann}}}, \bibinfo {author} {\bibfnamefont {F.}~\bibnamefont {{Dilssner}}}, \bibinfo {author} {\bibfnamefont {C.}~\bibnamefont {{Courde}}}, \bibinfo {author} {\bibfnamefont {S.}~\bibnamefont {{Bertone}}}, \bibinfo {author} {\bibfnamefont {F.}~\bibnamefont {{Gonzalez}}}, \bibinfo {author} {\bibfnamefont {A.}~\bibnamefont {{Hees}}}, \bibinfo {author} {\bibfnamefont {C.}~\bibnamefont {{Le Poncin-Lafitte}}}, \bibinfo {author} {\bibfnamefont {F.}~\bibnamefont {{Meynadier}}}, \bibinfo {author} {\bibfnamefont {R.}~\bibnamefont {{Prieto-Cerdeira}}}, \bibinfo {author} {\bibfnamefont {B.}~\bibnamefont {{Sohet}}}, \bibinfo {author} {\bibfnamefont {J.}~\bibnamefont {{Ventura-Traveset}}},\ and\ \bibinfo {author} {\bibfnamefont {P.}~\bibnamefont {{Wolf}}},\ }\bibfield  {title} {\bibinfo {title} {{Gravitational
  Redshift Test Using Eccentric Galileo Satellites}},\ }\href {https://doi.org/10.1103/PhysRevLett.121.231101} {\bibfield  {journal} {\bibinfo  {journal} {\prl}\ }\textbf {\bibinfo {volume} {121}},\ \bibinfo {eid} {231101} (\bibinfo {year} {2018})}\BibitemShut {NoStop}%
\bibitem [{\citenamefont {{Takamoto}}\ \emph {et~al.}(2020)\citenamefont {{Takamoto}}, \citenamefont {{Ushijima}}, \citenamefont {{Ohmae}}, \citenamefont {{Yahagi}}, \citenamefont {{Kokado}}, \citenamefont {{Shinkai}},\ and\ \citenamefont {{Katori}}}]{Takamoto_2020}%
  \BibitemOpen
  \bibfield  {author} {\bibinfo {author} {\bibfnamefont {M.}~\bibnamefont {{Takamoto}}}, \bibinfo {author} {\bibfnamefont {I.}~\bibnamefont {{Ushijima}}}, \bibinfo {author} {\bibfnamefont {N.}~\bibnamefont {{Ohmae}}}, \bibinfo {author} {\bibfnamefont {T.}~\bibnamefont {{Yahagi}}}, \bibinfo {author} {\bibfnamefont {K.}~\bibnamefont {{Kokado}}}, \bibinfo {author} {\bibfnamefont {H.}~\bibnamefont {{Shinkai}}},\ and\ \bibinfo {author} {\bibfnamefont {H.}~\bibnamefont {{Katori}}},\ }\bibfield  {title} {\bibinfo {title} {{Test of general relativity by a pair of transportable optical lattice clocks}},\ }\href {https://doi.org/10.1038/s41566-020-0619-8} {\bibfield  {journal} {\bibinfo  {journal} {Nature Photonics}\ }\textbf {\bibinfo {volume} {14}},\ \bibinfo {pages} {411} (\bibinfo {year} {2020})}\BibitemShut {NoStop}%
\bibitem [{\citenamefont {{Bothwell}}\ \emph {et~al.}(2022)\citenamefont {{Bothwell}}, \citenamefont {{Kennedy}}, \citenamefont {{Aeppli}}, \citenamefont {{Kedar}}, \citenamefont {{Robinson}}, \citenamefont {{Oelker}}, \citenamefont {{Staron}},\ and\ \citenamefont {{Ye}}}]{Bothwell_2022}%
  \BibitemOpen
  \bibfield  {author} {\bibinfo {author} {\bibfnamefont {T.}~\bibnamefont {{Bothwell}}}, \bibinfo {author} {\bibfnamefont {C.~J.}\ \bibnamefont {{Kennedy}}}, \bibinfo {author} {\bibfnamefont {A.}~\bibnamefont {{Aeppli}}}, \bibinfo {author} {\bibfnamefont {D.}~\bibnamefont {{Kedar}}}, \bibinfo {author} {\bibfnamefont {J.~M.}\ \bibnamefont {{Robinson}}}, \bibinfo {author} {\bibfnamefont {E.}~\bibnamefont {{Oelker}}}, \bibinfo {author} {\bibfnamefont {A.}~\bibnamefont {{Staron}}},\ and\ \bibinfo {author} {\bibfnamefont {J.}~\bibnamefont {{Ye}}},\ }\bibfield  {title} {\bibinfo {title} {{Resolving the gravitational redshift across a millimetre-scale atomic sample}},\ }\href {https://doi.org/10.1038/s41586-021-04349-7} {\bibfield  {journal} {\bibinfo  {journal} {Nature}\ }\textbf {\bibinfo {volume} {602}},\ \bibinfo {pages} {420} (\bibinfo {year} {2022})}\BibitemShut {NoStop}%
\bibitem [{\citenamefont {{Zheng}}\ \emph {et~al.}(2023)\citenamefont {{Zheng}}, \citenamefont {{Dolde}}, \citenamefont {{Cambria}}, \citenamefont {{Lim}},\ and\ \citenamefont {{Kolkowitz}}}]{zheng_2023}%
  \BibitemOpen
  \bibfield  {author} {\bibinfo {author} {\bibfnamefont {X.}~\bibnamefont {{Zheng}}}, \bibinfo {author} {\bibfnamefont {J.}~\bibnamefont {{Dolde}}}, \bibinfo {author} {\bibfnamefont {M.~C.}\ \bibnamefont {{Cambria}}}, \bibinfo {author} {\bibfnamefont {H.~M.}\ \bibnamefont {{Lim}}},\ and\ \bibinfo {author} {\bibfnamefont {S.}~\bibnamefont {{Kolkowitz}}},\ }\bibfield  {title} {\bibinfo {title} {{A lab-based test of the gravitational redshift with a miniature clock network}},\ }\href {https://doi.org/10.1038/s41467-023-40629-8} {\bibfield  {journal} {\bibinfo  {journal} {Nature Communications}\ }\textbf {\bibinfo {volume} {14}},\ \bibinfo {eid} {4886} (\bibinfo {year} {2023})}\BibitemShut {NoStop}%
\bibitem [{\citenamefont {{Bose}}\ \emph {et~al.}(2017)\citenamefont {{Bose}}, \citenamefont {{Mazumdar}}, \citenamefont {{Morley}}, \citenamefont {{Ulbricht}}, \citenamefont {{Toro{\v{s}}}}, \citenamefont {{Paternostro}}, \citenamefont {{Geraci}}, \citenamefont {{Barker}}, \citenamefont {{Kim}},\ and\ \citenamefont {{Milburn}}}]{Bose2017}%
  \BibitemOpen
  \bibfield  {author} {\bibinfo {author} {\bibfnamefont {S.}~\bibnamefont {{Bose}}}, \bibinfo {author} {\bibfnamefont {A.}~\bibnamefont {{Mazumdar}}}, \bibinfo {author} {\bibfnamefont {G.~W.}\ \bibnamefont {{Morley}}}, \bibinfo {author} {\bibfnamefont {H.}~\bibnamefont {{Ulbricht}}}, \bibinfo {author} {\bibfnamefont {M.}~\bibnamefont {{Toro{\v{s}}}}}, \bibinfo {author} {\bibfnamefont {M.}~\bibnamefont {{Paternostro}}}, \bibinfo {author} {\bibfnamefont {A.~A.}\ \bibnamefont {{Geraci}}}, \bibinfo {author} {\bibfnamefont {P.~F.}\ \bibnamefont {{Barker}}}, \bibinfo {author} {\bibfnamefont {M.~S.}\ \bibnamefont {{Kim}}},\ and\ \bibinfo {author} {\bibfnamefont {G.}~\bibnamefont {{Milburn}}},\ }\bibfield  {title} {\bibinfo {title} {{Spin Entanglement Witness for Quantum Gravity}},\ }\href {https://doi.org/10.1103/PhysRevLett.119.240401} {\bibfield  {journal} {\bibinfo  {journal} {\prl}\ }\textbf {\bibinfo {volume} {119}},\ \bibinfo {eid} {240401} (\bibinfo {year} {2017})}\BibitemShut {NoStop}%
\bibitem [{\citenamefont {{Marletto}}\ and\ \citenamefont {{Vedral}}(2017)}]{Marletto2017}%
  \BibitemOpen
  \bibfield  {author} {\bibinfo {author} {\bibfnamefont {C.}~\bibnamefont {{Marletto}}}\ and\ \bibinfo {author} {\bibfnamefont {V.}~\bibnamefont {{Vedral}}},\ }\bibfield  {title} {\bibinfo {title} {{Gravitationally Induced Entanglement between Two Massive Particles is Sufficient Evidence of Quantum Effects in Gravity}},\ }\href {https://doi.org/10.1103/PhysRevLett.119.240402} {\bibfield  {journal} {\bibinfo  {journal} {\prl}\ }\textbf {\bibinfo {volume} {119}},\ \bibinfo {eid} {240402} (\bibinfo {year} {2017})}\BibitemShut {NoStop}%
\bibitem [{\citenamefont {{Hall}}\ and\ \citenamefont {{Reginatto}}(2018)}]{Hall2018}%
  \BibitemOpen
  \bibfield  {author} {\bibinfo {author} {\bibfnamefont {M.~J.~W.}\ \bibnamefont {{Hall}}}\ and\ \bibinfo {author} {\bibfnamefont {M.}~\bibnamefont {{Reginatto}}},\ }\bibfield  {title} {\bibinfo {title} {{On two recent proposals for witnessing nonclassical gravity}},\ }\href {https://doi.org/10.1088/1751-8121/aaa734} {\bibfield  {journal} {\bibinfo  {journal} {Journal of Physics A Mathematical General}\ }\textbf {\bibinfo {volume} {51}},\ \bibinfo {eid} {085303} (\bibinfo {year} {2018})}\BibitemShut {NoStop}%
\bibitem [{\citenamefont {{Kent}}\ and\ \citenamefont {{Pital{\'u}a-Garc{\'\i}a}}(2021)}]{Kent2021}%
  \BibitemOpen
  \bibfield  {author} {\bibinfo {author} {\bibfnamefont {A.}~\bibnamefont {{Kent}}}\ and\ \bibinfo {author} {\bibfnamefont {D.}~\bibnamefont {{Pital{\'u}a-Garc{\'\i}a}}},\ }\bibfield  {title} {\bibinfo {title} {{Testing the nonclassicality of spacetime: What can we learn from Bell-Bose \textit{et al.}-Marletto-Vedral experiments?}},\ }\href {https://doi.org/10.1103/PhysRevD.104.126030} {\bibfield  {journal} {\bibinfo  {journal} {\prd}\ }\textbf {\bibinfo {volume} {104}},\ \bibinfo {eid} {126030} (\bibinfo {year} {2021})}\BibitemShut {NoStop}%
\bibitem [{\citenamefont {{Colella}}\ \emph {et~al.}(1975)\citenamefont {{Colella}}, \citenamefont {{Overhauser}},\ and\ \citenamefont {{Werner}}}]{COW_1975_PRL}%
  \BibitemOpen
  \bibfield  {author} {\bibinfo {author} {\bibfnamefont {R.}~\bibnamefont {{Colella}}}, \bibinfo {author} {\bibfnamefont {A.~W.}\ \bibnamefont {{Overhauser}}},\ and\ \bibinfo {author} {\bibfnamefont {S.~A.}\ \bibnamefont {{Werner}}},\ }\bibfield  {title} {\bibinfo {title} {{Observation of Gravitationally Induced Quantum Interference}},\ }\href {https://doi.org/10.1103/PhysRevLett.34.1472} {\bibfield  {journal} {\bibinfo  {journal} {\prl}\ }\textbf {\bibinfo {volume} {34}},\ \bibinfo {pages} {1472} (\bibinfo {year} {1975})}\BibitemShut {NoStop}%
\bibitem [{\citenamefont {{Kasevich}}\ and\ \citenamefont {{Chu}}(1991)}]{Kasevich1991}%
  \BibitemOpen
  \bibfield  {author} {\bibinfo {author} {\bibfnamefont {M.}~\bibnamefont {{Kasevich}}}\ and\ \bibinfo {author} {\bibfnamefont {S.}~\bibnamefont {{Chu}}},\ }\bibfield  {title} {\bibinfo {title} {{Atomic interferometry using stimulated Raman transitions}},\ }\href {https://doi.org/10.1103/PhysRevLett.67.181} {\bibfield  {journal} {\bibinfo  {journal} {\prl}\ }\textbf {\bibinfo {volume} {67}},\ \bibinfo {pages} {181} (\bibinfo {year} {1991})}\BibitemShut {NoStop}%
\bibitem [{\citenamefont {{Kasevich}}\ and\ \citenamefont {{Chu}}(1992)}]{Kasevich1992}%
  \BibitemOpen
  \bibfield  {author} {\bibinfo {author} {\bibfnamefont {M.}~\bibnamefont {{Kasevich}}}\ and\ \bibinfo {author} {\bibfnamefont {S.}~\bibnamefont {{Chu}}},\ }\bibfield  {title} {\bibinfo {title} {{Measurement of the gravitational acceleration of an atom with a light-pulse atom interferometer}},\ }\href {https://doi.org/10.1007/BF00325375} {\bibfield  {journal} {\bibinfo  {journal} {Applied Physics B: Lasers and Optics}\ }\textbf {\bibinfo {volume} {54}},\ \bibinfo {pages} {321} (\bibinfo {year} {1992})}\BibitemShut {NoStop}%
\bibitem [{\citenamefont {{Snadden}}\ \emph {et~al.}(1998)\citenamefont {{Snadden}}, \citenamefont {{McGuirk}}, \citenamefont {{Bouyer}}, \citenamefont {{Haritos}},\ and\ \citenamefont {{Kasevich}}}]{Snadden1998}%
  \BibitemOpen
  \bibfield  {author} {\bibinfo {author} {\bibfnamefont {M.~J.}\ \bibnamefont {{Snadden}}}, \bibinfo {author} {\bibfnamefont {J.~M.}\ \bibnamefont {{McGuirk}}}, \bibinfo {author} {\bibfnamefont {P.}~\bibnamefont {{Bouyer}}}, \bibinfo {author} {\bibfnamefont {K.~G.}\ \bibnamefont {{Haritos}}},\ and\ \bibinfo {author} {\bibfnamefont {M.~A.}\ \bibnamefont {{Kasevich}}},\ }\bibfield  {title} {\bibinfo {title} {{Measurement of the Earth's Gravity Gradient with an Atom Interferometer-Based Gravity Gradiometer}},\ }\href {https://doi.org/10.1103/PhysRevLett.81.971} {\bibfield  {journal} {\bibinfo  {journal} {\prl}\ }\textbf {\bibinfo {volume} {81}},\ \bibinfo {pages} {971} (\bibinfo {year} {1998})}\BibitemShut {NoStop}%
\bibitem [{\citenamefont {{Altin}}\ \emph {et~al.}(2013)\citenamefont {{Altin}}, \citenamefont {{Johnsson}}, \citenamefont {{Negnevitsky}}, \citenamefont {{Dennis}}, \citenamefont {{Anderson}}, \citenamefont {{Debs}}, \citenamefont {{Szigeti}}, \citenamefont {{Hardman}}, \citenamefont {{Bennetts}}, \citenamefont {{McDonald}}, \citenamefont {{Turner}}, \citenamefont {{Close}},\ and\ \citenamefont {{Robins}}}]{Altin2013}%
  \BibitemOpen
  \bibfield  {author} {\bibinfo {author} {\bibfnamefont {P.~A.}\ \bibnamefont {{Altin}}}, \bibinfo {author} {\bibfnamefont {M.~T.}\ \bibnamefont {{Johnsson}}}, \bibinfo {author} {\bibfnamefont {V.}~\bibnamefont {{Negnevitsky}}}, \bibinfo {author} {\bibfnamefont {G.~R.}\ \bibnamefont {{Dennis}}}, \bibinfo {author} {\bibfnamefont {R.~P.}\ \bibnamefont {{Anderson}}}, \bibinfo {author} {\bibfnamefont {J.~E.}\ \bibnamefont {{Debs}}}, \bibinfo {author} {\bibfnamefont {S.~S.}\ \bibnamefont {{Szigeti}}}, \bibinfo {author} {\bibfnamefont {K.~S.}\ \bibnamefont {{Hardman}}}, \bibinfo {author} {\bibfnamefont {S.}~\bibnamefont {{Bennetts}}}, \bibinfo {author} {\bibfnamefont {G.~D.}\ \bibnamefont {{McDonald}}}, \bibinfo {author} {\bibfnamefont {L.~D.}\ \bibnamefont {{Turner}}}, \bibinfo {author} {\bibfnamefont {J.~D.}\ \bibnamefont {{Close}}},\ and\ \bibinfo {author} {\bibfnamefont {N.~P.}\ \bibnamefont {{Robins}}},\ }\bibfield  {title} {\bibinfo {title} {{Precision atomic gravimeter based on Bragg diffraction}},\ }\href
  {https://doi.org/10.1088/1367-2630/15/2/023009} {\bibfield  {journal} {\bibinfo  {journal} {New Journal of Physics}\ }\textbf {\bibinfo {volume} {15}},\ \bibinfo {eid} {023009} (\bibinfo {year} {2013})}\BibitemShut {NoStop}%
\bibitem [{\citenamefont {{Xu}}\ \emph {et~al.}(2019)\citenamefont {{Xu}}, \citenamefont {{Jaffe}}, \citenamefont {{Panda}}, \citenamefont {{Kristensen}}, \citenamefont {{Clark}},\ and\ \citenamefont {{M{\"u}ller}}}]{Xu2019}%
  \BibitemOpen
  \bibfield  {author} {\bibinfo {author} {\bibfnamefont {V.}~\bibnamefont {{Xu}}}, \bibinfo {author} {\bibfnamefont {M.}~\bibnamefont {{Jaffe}}}, \bibinfo {author} {\bibfnamefont {C.~D.}\ \bibnamefont {{Panda}}}, \bibinfo {author} {\bibfnamefont {S.~L.}\ \bibnamefont {{Kristensen}}}, \bibinfo {author} {\bibfnamefont {L.~W.}\ \bibnamefont {{Clark}}},\ and\ \bibinfo {author} {\bibfnamefont {H.}~\bibnamefont {{M{\"u}ller}}},\ }\bibfield  {title} {\bibinfo {title} {{Probing gravity by holding atoms for 20 seconds}},\ }\href {https://doi.org/10.1126/science.aay6428} {\bibfield  {journal} {\bibinfo  {journal} {Science}\ }\textbf {\bibinfo {volume} {366}},\ \bibinfo {pages} {745} (\bibinfo {year} {2019})}\BibitemShut {NoStop}%
\bibitem [{\citenamefont {{Panda}}\ \emph {et~al.}(2024)\citenamefont {{Panda}}, \citenamefont {{Tao}}, \citenamefont {{Ceja}}, \citenamefont {{Khoury}}, \citenamefont {{Tino}},\ and\ \citenamefont {{M{\"u}ller}}}]{Panda2024}%
  \BibitemOpen
  \bibfield  {author} {\bibinfo {author} {\bibfnamefont {C.~D.}\ \bibnamefont {{Panda}}}, \bibinfo {author} {\bibfnamefont {M.~J.}\ \bibnamefont {{Tao}}}, \bibinfo {author} {\bibfnamefont {M.}~\bibnamefont {{Ceja}}}, \bibinfo {author} {\bibfnamefont {J.}~\bibnamefont {{Khoury}}}, \bibinfo {author} {\bibfnamefont {G.~M.}\ \bibnamefont {{Tino}}},\ and\ \bibinfo {author} {\bibfnamefont {H.}~\bibnamefont {{M{\"u}ller}}},\ }\bibfield  {title} {\bibinfo {title} {{Measuring gravitational attraction with a lattice atom interferometer}},\ }\href {https://doi.org/10.1038/s41586-024-07561-3} {\bibfield  {journal} {\bibinfo  {journal} {\nat}\ }\textbf {\bibinfo {volume} {631}},\ \bibinfo {pages} {515} (\bibinfo {year} {2024})}\BibitemShut {NoStop}%
\bibitem [{\citenamefont {{Overstreet}}\ \emph {et~al.}(2022)\citenamefont {{Overstreet}}, \citenamefont {{Asenbaum}}, \citenamefont {{Curti}}, \citenamefont {{Kim}},\ and\ \citenamefont {{Kasevich}}}]{Overstreet2022}%
  \BibitemOpen
  \bibfield  {author} {\bibinfo {author} {\bibfnamefont {C.}~\bibnamefont {{Overstreet}}}, \bibinfo {author} {\bibfnamefont {P.}~\bibnamefont {{Asenbaum}}}, \bibinfo {author} {\bibfnamefont {J.}~\bibnamefont {{Curti}}}, \bibinfo {author} {\bibfnamefont {M.}~\bibnamefont {{Kim}}},\ and\ \bibinfo {author} {\bibfnamefont {M.~A.}\ \bibnamefont {{Kasevich}}},\ }\bibfield  {title} {\bibinfo {title} {{Observation of a gravitational Aharonov-Bohm effect}},\ }\href {https://doi.org/10.1126/science.abl7152} {\bibfield  {journal} {\bibinfo  {journal} {Science}\ }\textbf {\bibinfo {volume} {375}},\ \bibinfo {pages} {226} (\bibinfo {year} {2022})}\BibitemShut {NoStop}%
\bibitem [{\citenamefont {{Zych}}\ \emph {et~al.}(2012)\citenamefont {{Zych}}, \citenamefont {{Costa}}, \citenamefont {{Pikovski}}, \citenamefont {{Ralph}},\ and\ \citenamefont {{Brukner}}}]{Zych_2012}%
  \BibitemOpen
  \bibfield  {author} {\bibinfo {author} {\bibfnamefont {M.}~\bibnamefont {{Zych}}}, \bibinfo {author} {\bibfnamefont {F.}~\bibnamefont {{Costa}}}, \bibinfo {author} {\bibfnamefont {I.}~\bibnamefont {{Pikovski}}}, \bibinfo {author} {\bibfnamefont {T.~C.}\ \bibnamefont {{Ralph}}},\ and\ \bibinfo {author} {\bibfnamefont {{\v{C}}.}~\bibnamefont {{Brukner}}},\ }\bibfield  {title} {\bibinfo {title} {{General relativistic effects in quantum interference of photons}},\ }\href {https://doi.org/10.1088/0264-9381/29/22/224010} {\bibfield  {journal} {\bibinfo  {journal} {Classical and Quantum Gravity}\ }\textbf {\bibinfo {volume} {29}},\ \bibinfo {eid} {224010} (\bibinfo {year} {2012})}\BibitemShut {NoStop}%
\bibitem [{\citenamefont {{Mieling}}\ \emph {et~al.}(2025)\citenamefont {{Mieling}}, \citenamefont {{Morling}}, \citenamefont {{Hilweg}},\ and\ \citenamefont {{Walther}}}]{MMHW_2025}%
  \BibitemOpen
  \bibfield  {author} {\bibinfo {author} {\bibfnamefont {T.~B.}\ \bibnamefont {{Mieling}}}, \bibinfo {author} {\bibfnamefont {T.}~\bibnamefont {{Morling}}}, \bibinfo {author} {\bibfnamefont {C.}~\bibnamefont {{Hilweg}}},\ and\ \bibinfo {author} {\bibfnamefont {P.}~\bibnamefont {{Walther}}},\ }\href@noop {} {\bibinfo {title} {Quantum interferometry in external gravitational fields}} (\bibinfo {year} {2025}),\ \Eprint {https://arxiv.org/abs/2507.21808} {arXiv:2507.21808 [gr-qc]} \BibitemShut {NoStop}%
\bibitem [{\citenamefont {Rideout}\ \emph {et~al.}(2012)\citenamefont {Rideout}, \citenamefont {Jennewein}, \citenamefont {Amelino-Camelia}, \citenamefont {Demarie}, \citenamefont {Higgins}, \citenamefont {Kempf}, \citenamefont {Kent}, \citenamefont {Laflamme}, \citenamefont {Ma}, \citenamefont {Mann}, \citenamefont {Martín-Martínez}, \citenamefont {Menicucci}, \citenamefont {Moffat}, \citenamefont {Simon}, \citenamefont {Sorkin}, \citenamefont {Smolin},\ and\ \citenamefont {Terno}}]{Rideout2012}%
  \BibitemOpen
  \bibfield  {author} {\bibinfo {author} {\bibfnamefont {D.}~\bibnamefont {Rideout}}, \bibinfo {author} {\bibfnamefont {T.}~\bibnamefont {Jennewein}}, \bibinfo {author} {\bibfnamefont {G.}~\bibnamefont {Amelino-Camelia}}, \bibinfo {author} {\bibfnamefont {T.~F.}\ \bibnamefont {Demarie}}, \bibinfo {author} {\bibfnamefont {B.~L.}\ \bibnamefont {Higgins}}, \bibinfo {author} {\bibfnamefont {A.}~\bibnamefont {Kempf}}, \bibinfo {author} {\bibfnamefont {A.}~\bibnamefont {Kent}}, \bibinfo {author} {\bibfnamefont {R.}~\bibnamefont {Laflamme}}, \bibinfo {author} {\bibfnamefont {X.}~\bibnamefont {Ma}}, \bibinfo {author} {\bibfnamefont {R.~B.}\ \bibnamefont {Mann}}, \bibinfo {author} {\bibfnamefont {E.}~\bibnamefont {Martín-Martínez}}, \bibinfo {author} {\bibfnamefont {N.~C.}\ \bibnamefont {Menicucci}}, \bibinfo {author} {\bibfnamefont {J.}~\bibnamefont {Moffat}}, \bibinfo {author} {\bibfnamefont {C.}~\bibnamefont {Simon}}, \bibinfo {author} {\bibfnamefont {R.}~\bibnamefont {Sorkin}}, \bibinfo {author} {\bibfnamefont
  {L.}~\bibnamefont {Smolin}},\ and\ \bibinfo {author} {\bibfnamefont {D.~R.}\ \bibnamefont {Terno}},\ }\bibfield  {title} {\bibinfo {title} {Fundamental quantum optics experiments conceivable with satellites—reaching relativistic distances and velocities},\ }\href {https://doi.org/10.1088/0264-9381/29/22/224011} {\bibfield  {journal} {\bibinfo  {journal} {Classical and Quantum Gravity}\ }\textbf {\bibinfo {volume} {29}},\ \bibinfo {pages} {224011} (\bibinfo {year} {2012})}\BibitemShut {NoStop}%
\bibitem [{\citenamefont {{Vallone}}\ \emph {et~al.}(2016)\citenamefont {{Vallone}}, \citenamefont {{Dequal}}, \citenamefont {{Tomasin}}, \citenamefont {{Vedovato}}, \citenamefont {{Schiavon}}, \citenamefont {{Luceri}}, \citenamefont {{Bianco}},\ and\ \citenamefont {{Villoresi}}}]{vallone_2016}%
  \BibitemOpen
  \bibfield  {author} {\bibinfo {author} {\bibfnamefont {G.}~\bibnamefont {{Vallone}}}, \bibinfo {author} {\bibfnamefont {D.}~\bibnamefont {{Dequal}}}, \bibinfo {author} {\bibfnamefont {M.}~\bibnamefont {{Tomasin}}}, \bibinfo {author} {\bibfnamefont {F.}~\bibnamefont {{Vedovato}}}, \bibinfo {author} {\bibfnamefont {M.}~\bibnamefont {{Schiavon}}}, \bibinfo {author} {\bibfnamefont {V.}~\bibnamefont {{Luceri}}}, \bibinfo {author} {\bibfnamefont {G.}~\bibnamefont {{Bianco}}},\ and\ \bibinfo {author} {\bibfnamefont {P.}~\bibnamefont {{Villoresi}}},\ }\bibfield  {title} {\bibinfo {title} {{Interference at the Single Photon Level Along Satellite-Ground Channels}},\ }\href {https://doi.org/10.1103/PhysRevLett.116.253601} {\bibfield  {journal} {\bibinfo  {journal} {\prl}\ }\textbf {\bibinfo {volume} {116}},\ \bibinfo {eid} {253601} (\bibinfo {year} {2016})}\BibitemShut {NoStop}%
\bibitem [{\citenamefont {Xu}\ \emph {et~al.}(2019)\citenamefont {Xu}, \citenamefont {Ma}, \citenamefont {Ren}, \citenamefont {Yong}, \citenamefont {Ralph}, \citenamefont {Liao}, \citenamefont {Yin}, \citenamefont {Liu}, \citenamefont {Cai}, \citenamefont {Han}, \citenamefont {Wu}, \citenamefont {Wang}, \citenamefont {Li}, \citenamefont {Yang}, \citenamefont {Lin}, \citenamefont {Li}, \citenamefont {Liu}, \citenamefont {Chen}, \citenamefont {Lu}, \citenamefont {Chen}, \citenamefont {Fan}, \citenamefont {Peng},\ and\ \citenamefont {Pan}}]{Ping_2019}%
  \BibitemOpen
  \bibfield  {author} {\bibinfo {author} {\bibfnamefont {P.}~\bibnamefont {Xu}}, \bibinfo {author} {\bibfnamefont {Y.}~\bibnamefont {Ma}}, \bibinfo {author} {\bibfnamefont {J.-G.}\ \bibnamefont {Ren}}, \bibinfo {author} {\bibfnamefont {H.-L.}\ \bibnamefont {Yong}}, \bibinfo {author} {\bibfnamefont {T.~C.}\ \bibnamefont {Ralph}}, \bibinfo {author} {\bibfnamefont {S.-K.}\ \bibnamefont {Liao}}, \bibinfo {author} {\bibfnamefont {J.}~\bibnamefont {Yin}}, \bibinfo {author} {\bibfnamefont {W.-Y.}\ \bibnamefont {Liu}}, \bibinfo {author} {\bibfnamefont {W.-Q.}\ \bibnamefont {Cai}}, \bibinfo {author} {\bibfnamefont {X.}~\bibnamefont {Han}}, \bibinfo {author} {\bibfnamefont {H.-N.}\ \bibnamefont {Wu}}, \bibinfo {author} {\bibfnamefont {W.-Y.}\ \bibnamefont {Wang}}, \bibinfo {author} {\bibfnamefont {F.-Z.}\ \bibnamefont {Li}}, \bibinfo {author} {\bibfnamefont {M.}~\bibnamefont {Yang}}, \bibinfo {author} {\bibfnamefont {F.-L.}\ \bibnamefont {Lin}}, \bibinfo {author} {\bibfnamefont {L.}~\bibnamefont {Li}}, \bibinfo
  {author} {\bibfnamefont {N.-L.}\ \bibnamefont {Liu}}, \bibinfo {author} {\bibfnamefont {Y.-A.}\ \bibnamefont {Chen}}, \bibinfo {author} {\bibfnamefont {C.-Y.}\ \bibnamefont {Lu}}, \bibinfo {author} {\bibfnamefont {Y.}~\bibnamefont {Chen}}, \bibinfo {author} {\bibfnamefont {J.}~\bibnamefont {Fan}}, \bibinfo {author} {\bibfnamefont {C.-Z.}\ \bibnamefont {Peng}},\ and\ \bibinfo {author} {\bibfnamefont {J.-W.}\ \bibnamefont {Pan}},\ }\bibfield  {title} {\bibinfo {title} {{Satellite testing of a gravitationally induced quantum decoherence model}},\ }\href {https://doi.org/10.1126/science.aay5820} {\bibfield  {journal} {\bibinfo  {journal} {Science}\ }\textbf {\bibinfo {volume} {366}},\ \bibinfo {pages} {132} (\bibinfo {year} {2019})}\BibitemShut {NoStop}%
\bibitem [{\citenamefont {Lu}\ \emph {et~al.}(2022)\citenamefont {Lu}, \citenamefont {Cao}, \citenamefont {Peng},\ and\ \citenamefont {Pan}}]{RMP_Micius_2022}%
  \BibitemOpen
  \bibfield  {author} {\bibinfo {author} {\bibfnamefont {C.-Y.}\ \bibnamefont {Lu}}, \bibinfo {author} {\bibfnamefont {Y.}~\bibnamefont {Cao}}, \bibinfo {author} {\bibfnamefont {C.-Z.}\ \bibnamefont {Peng}},\ and\ \bibinfo {author} {\bibfnamefont {J.-W.}\ \bibnamefont {Pan}},\ }\bibfield  {title} {\bibinfo {title} {{Micius quantum experiments in space}},\ }\href {https://doi.org/10.1103/RevModPhys.94.035001} {\bibfield  {journal} {\bibinfo  {journal} {Reviews of Modern Physics}\ }\textbf {\bibinfo {volume} {94}},\ \bibinfo {eid} {035001} (\bibinfo {year} {2022})}\BibitemShut {NoStop}%
\bibitem [{\citenamefont {{Barzel}}\ \emph {et~al.}(2022)\citenamefont {{Barzel}}, \citenamefont {{Bruschi}}, \citenamefont {{Schell}},\ and\ \citenamefont {{L{\"a}mmerzahl}}}]{2022PhRvD.105j5016B}%
  \BibitemOpen
  \bibfield  {author} {\bibinfo {author} {\bibfnamefont {R.}~\bibnamefont {{Barzel}}}, \bibinfo {author} {\bibfnamefont {D.~E.}\ \bibnamefont {{Bruschi}}}, \bibinfo {author} {\bibfnamefont {A.~W.}\ \bibnamefont {{Schell}}},\ and\ \bibinfo {author} {\bibfnamefont {C.}~\bibnamefont {{L{\"a}mmerzahl}}},\ }\bibfield  {title} {\bibinfo {title} {{Observer dependence of photon bunching: The influence of the relativistic redshift on Hong-Ou-Mandel interference}},\ }\href {https://doi.org/10.1103/PhysRevD.105.105016} {\bibfield  {journal} {\bibinfo  {journal} {\prd}\ }\textbf {\bibinfo {volume} {105}},\ \bibinfo {eid} {105016} (\bibinfo {year} {2022})}\BibitemShut {NoStop}%
\bibitem [{\citenamefont {Mieling}\ \emph {et~al.}(2022)\citenamefont {Mieling}, \citenamefont {Hilweg},\ and\ \citenamefont {Walther}}]{Mieling2022}%
  \BibitemOpen
  \bibfield  {author} {\bibinfo {author} {\bibfnamefont {T.~B.}\ \bibnamefont {Mieling}}, \bibinfo {author} {\bibfnamefont {C.}~\bibnamefont {Hilweg}},\ and\ \bibinfo {author} {\bibfnamefont {P.}~\bibnamefont {Walther}},\ }\bibfield  {title} {\bibinfo {title} {Measuring space-time curvature using maximally path-entangled quantum states},\ }\href {https://doi.org/10.1103/PhysRevA.106.L031701} {\bibfield  {journal} {\bibinfo  {journal} {Phys. Rev. A}\ }\textbf {\bibinfo {volume} {106}},\ \bibinfo {pages} {L031701} (\bibinfo {year} {2022})}\BibitemShut {NoStop}%
\bibitem [{\citenamefont {Mohageg}\ \emph {et~al.}(2022)\citenamefont {Mohageg}, \citenamefont {Mazzarella}, \citenamefont {Anastopoulos}, \citenamefont {Gallicchio}, \citenamefont {Hu}, \citenamefont {Jennewein}, \citenamefont {Johnson}, \citenamefont {Lin}, \citenamefont {Ling}, \citenamefont {Marquardt}, \citenamefont {Meister}, \citenamefont {Newell}, \citenamefont {Roura}, \citenamefont {Schleich}, \citenamefont {Schubert}, \citenamefont {Strekalov}, \citenamefont {Vallone}, \citenamefont {Villoresi}, \citenamefont {Wörner}, \citenamefont {Yu}, \citenamefont {Zhai},\ and\ \citenamefont {Kwiat}}]{Mohageg2022}%
  \BibitemOpen
  \bibfield  {author} {\bibinfo {author} {\bibfnamefont {M.}~\bibnamefont {Mohageg}}, \bibinfo {author} {\bibfnamefont {L.}~\bibnamefont {Mazzarella}}, \bibinfo {author} {\bibfnamefont {C.}~\bibnamefont {Anastopoulos}}, \bibinfo {author} {\bibfnamefont {J.}~\bibnamefont {Gallicchio}}, \bibinfo {author} {\bibfnamefont {B.-L.}\ \bibnamefont {Hu}}, \bibinfo {author} {\bibfnamefont {T.}~\bibnamefont {Jennewein}}, \bibinfo {author} {\bibfnamefont {S.}~\bibnamefont {Johnson}}, \bibinfo {author} {\bibfnamefont {S.-Y.}\ \bibnamefont {Lin}}, \bibinfo {author} {\bibfnamefont {A.}~\bibnamefont {Ling}}, \bibinfo {author} {\bibfnamefont {C.}~\bibnamefont {Marquardt}}, \bibinfo {author} {\bibfnamefont {M.}~\bibnamefont {Meister}}, \bibinfo {author} {\bibfnamefont {R.}~\bibnamefont {Newell}}, \bibinfo {author} {\bibfnamefont {A.}~\bibnamefont {Roura}}, \bibinfo {author} {\bibfnamefont {W.~P.}\ \bibnamefont {Schleich}}, \bibinfo {author} {\bibfnamefont {C.}~\bibnamefont {Schubert}}, \bibinfo {author} {\bibfnamefont {D.~V.}\
  \bibnamefont {Strekalov}}, \bibinfo {author} {\bibfnamefont {G.}~\bibnamefont {Vallone}}, \bibinfo {author} {\bibfnamefont {P.}~\bibnamefont {Villoresi}}, \bibinfo {author} {\bibfnamefont {L.}~\bibnamefont {Wörner}}, \bibinfo {author} {\bibfnamefont {N.}~\bibnamefont {Yu}}, \bibinfo {author} {\bibfnamefont {A.}~\bibnamefont {Zhai}},\ and\ \bibinfo {author} {\bibfnamefont {P.}~\bibnamefont {Kwiat}},\ }\bibfield  {title} {\bibinfo {title} {The deep space quantum link: prospective fundamental physics experiments using long-baseline quantum optics},\ }\href {https://doi.org/10.1140/epjqt/s40507-022-00143-0} {\bibfield  {journal} {\bibinfo  {journal} {EPJ Quantum Technology}\ }\textbf {\bibinfo {volume} {9}},\ \bibinfo {pages} {25} (\bibinfo {year} {2022})}\BibitemShut {NoStop}%
\bibitem [{\citenamefont {{Wu}}\ \emph {et~al.}(2024)\citenamefont {{Wu}}, \citenamefont {{Li}}, \citenamefont {{Li}}, \citenamefont {{You}}, \citenamefont {{Liu}}, \citenamefont {{Ren}}, \citenamefont {{Yin}}, \citenamefont {{Lu}}, \citenamefont {{Cao}}, \citenamefont {{Peng}},\ and\ \citenamefont {{Pan}}}]{Wu_2024_PRL}%
  \BibitemOpen
  \bibfield  {author} {\bibinfo {author} {\bibfnamefont {H.-N.}\ \bibnamefont {{Wu}}}, \bibinfo {author} {\bibfnamefont {Y.-H.}\ \bibnamefont {{Li}}}, \bibinfo {author} {\bibfnamefont {B.}~\bibnamefont {{Li}}}, \bibinfo {author} {\bibfnamefont {X.}~\bibnamefont {{You}}}, \bibinfo {author} {\bibfnamefont {R.-Z.}\ \bibnamefont {{Liu}}}, \bibinfo {author} {\bibfnamefont {J.-G.}\ \bibnamefont {{Ren}}}, \bibinfo {author} {\bibfnamefont {J.}~\bibnamefont {{Yin}}}, \bibinfo {author} {\bibfnamefont {C.-Y.}\ \bibnamefont {{Lu}}}, \bibinfo {author} {\bibfnamefont {Y.}~\bibnamefont {{Cao}}}, \bibinfo {author} {\bibfnamefont {C.-Z.}\ \bibnamefont {{Peng}}},\ and\ \bibinfo {author} {\bibfnamefont {J.-W.}\ \bibnamefont {{Pan}}},\ }\bibfield  {title} {\bibinfo {title} {{Single-Photon Interference over 8.4 km Urban Atmosphere: Toward Testing Quantum Effects in Curved Spacetime with Photons}},\ }\href {https://doi.org/10.1103/PhysRevLett.133.020201} {\bibfield  {journal} {\bibinfo  {journal} {\prl}\ }\textbf {\bibinfo
  {volume} {133}},\ \bibinfo {eid} {020201} (\bibinfo {year} {2024})}\BibitemShut {NoStop}%
\bibitem [{\citenamefont {Stodolsky}(1979)}]{Stodolsky1979}%
  \BibitemOpen
  \bibfield  {author} {\bibinfo {author} {\bibfnamefont {L.}~\bibnamefont {Stodolsky}},\ }\bibfield  {title} {\bibinfo {title} {Matter and light wave interferometry in gravitational fields},\ }\href {https://doi.org/10.1007/BF00759302} {\bibfield  {journal} {\bibinfo  {journal} {General Relativity and Gravitation}\ }\textbf {\bibinfo {volume} {11}},\ \bibinfo {pages} {391} (\bibinfo {year} {1979})}\BibitemShut {NoStop}%
\bibitem [{\citenamefont {Tanaka}(1983)}]{Tanaka1983}%
  \BibitemOpen
  \bibfield  {author} {\bibinfo {author} {\bibfnamefont {K.}~\bibnamefont {Tanaka}},\ }\bibfield  {title} {\bibinfo {title} {How to detect the gravitationally induced phase shift of electromagnetic waves by optical-fiber interferometry},\ }\href {https://doi.org/10.1103/PhysRevLett.51.378} {\bibfield  {journal} {\bibinfo  {journal} {Phys. Rev. Lett.}\ }\textbf {\bibinfo {volume} {51}},\ \bibinfo {pages} {378} (\bibinfo {year} {1983})}\BibitemShut {NoStop}%
\bibitem [{\citenamefont {{Hilweg}}\ \emph {et~al.}(2017)\citenamefont {{Hilweg}}, \citenamefont {{Massa}}, \citenamefont {{Martynov}}, \citenamefont {{Mavalvala}}, \citenamefont {{Chru{\'s}ciel}},\ and\ \citenamefont {{Walther}}}]{Hilweg_2017}%
  \BibitemOpen
  \bibfield  {author} {\bibinfo {author} {\bibfnamefont {C.}~\bibnamefont {{Hilweg}}}, \bibinfo {author} {\bibfnamefont {F.}~\bibnamefont {{Massa}}}, \bibinfo {author} {\bibfnamefont {D.}~\bibnamefont {{Martynov}}}, \bibinfo {author} {\bibfnamefont {N.}~\bibnamefont {{Mavalvala}}}, \bibinfo {author} {\bibfnamefont {P.~T.}\ \bibnamefont {{Chru{\'s}ciel}}},\ and\ \bibinfo {author} {\bibfnamefont {P.}~\bibnamefont {{Walther}}},\ }\bibfield  {title} {\bibinfo {title} {{Gravitationally induced phase shift on a single photon}},\ }\href {https://doi.org/10.1088/1367-2630/aa638f} {\bibfield  {journal} {\bibinfo  {journal} {New Journal of Physics}\ }\textbf {\bibinfo {volume} {19}},\ \bibinfo {eid} {033028} (\bibinfo {year} {2017})}\BibitemShut {NoStop}%
\bibitem [{\citenamefont {Chen}\ and\ \citenamefont {Ralph}(2019)}]{Chen2019}%
  \BibitemOpen
  \bibfield  {author} {\bibinfo {author} {\bibfnamefont {S.~Y.}\ \bibnamefont {Chen}}\ and\ \bibinfo {author} {\bibfnamefont {T.~C.}\ \bibnamefont {Ralph}},\ }\bibfield  {title} {\bibinfo {title} {Estimation of gravitational acceleration with quantum optical interferometers},\ }\href {https://doi.org/10.1103/PhysRevA.99.023803} {\bibfield  {journal} {\bibinfo  {journal} {Phys. Rev. A}\ }\textbf {\bibinfo {volume} {99}},\ \bibinfo {pages} {023803} (\bibinfo {year} {2019})}\BibitemShut {NoStop}%
\bibitem [{\citenamefont {Barzel}\ \emph {et~al.}(2024)\citenamefont {Barzel}, \citenamefont {G{\"{u}}ndo{\u{g}}an}, \citenamefont {Krutzik}, \citenamefont {R{\"{a}}tzel},\ and\ \citenamefont {L{\"{a}}mmerzahl}}]{Barzel2024}%
  \BibitemOpen
  \bibfield  {author} {\bibinfo {author} {\bibfnamefont {R.}~\bibnamefont {Barzel}}, \bibinfo {author} {\bibfnamefont {M.}~\bibnamefont {G{\"{u}}ndo{\u{g}}an}}, \bibinfo {author} {\bibfnamefont {M.}~\bibnamefont {Krutzik}}, \bibinfo {author} {\bibfnamefont {D.}~\bibnamefont {R{\"{a}}tzel}},\ and\ \bibinfo {author} {\bibfnamefont {C.}~\bibnamefont {L{\"{a}}mmerzahl}},\ }\bibfield  {title} {\bibinfo {title} {Entanglement dynamics of photon pairs and quantum memories in the gravitational field of the earth},\ }\href {https://doi.org/10.22331/q-2024-02-29-1273} {\bibfield  {journal} {\bibinfo  {journal} {{Quantum}}\ }\textbf {\bibinfo {volume} {8}},\ \bibinfo {pages} {1273} (\bibinfo {year} {2024})}\BibitemShut {NoStop}%
\bibitem [{\citenamefont {Gisin}\ \emph {et~al.}(2002)\citenamefont {Gisin}, \citenamefont {Ribordy}, \citenamefont {Tittel},\ and\ \citenamefont {Zbinden}}]{RevModPhys.74.145}%
  \BibitemOpen
  \bibfield  {author} {\bibinfo {author} {\bibfnamefont {N.}~\bibnamefont {Gisin}}, \bibinfo {author} {\bibfnamefont {G.}~\bibnamefont {Ribordy}}, \bibinfo {author} {\bibfnamefont {W.}~\bibnamefont {Tittel}},\ and\ \bibinfo {author} {\bibfnamefont {H.}~\bibnamefont {Zbinden}},\ }\bibfield  {title} {\bibinfo {title} {Quantum cryptography},\ }\href {https://doi.org/10.1103/RevModPhys.74.145} {\bibfield  {journal} {\bibinfo  {journal} {Rev. Mod. Phys.}\ }\textbf {\bibinfo {volume} {74}},\ \bibinfo {pages} {145} (\bibinfo {year} {2002})}\BibitemShut {NoStop}%
\bibitem [{\citenamefont {Kok}\ \emph {et~al.}(2007)\citenamefont {Kok}, \citenamefont {Munro}, \citenamefont {Nemoto}, \citenamefont {Ralph}, \citenamefont {Dowling},\ and\ \citenamefont {Milburn}}]{RevModPhys.79.135}%
  \BibitemOpen
  \bibfield  {author} {\bibinfo {author} {\bibfnamefont {P.}~\bibnamefont {Kok}}, \bibinfo {author} {\bibfnamefont {W.~J.}\ \bibnamefont {Munro}}, \bibinfo {author} {\bibfnamefont {K.}~\bibnamefont {Nemoto}}, \bibinfo {author} {\bibfnamefont {T.~C.}\ \bibnamefont {Ralph}}, \bibinfo {author} {\bibfnamefont {J.~P.}\ \bibnamefont {Dowling}},\ and\ \bibinfo {author} {\bibfnamefont {G.~J.}\ \bibnamefont {Milburn}},\ }\bibfield  {title} {\bibinfo {title} {Linear optical quantum computing with photonic qubits},\ }\href {https://doi.org/10.1103/RevModPhys.79.135} {\bibfield  {journal} {\bibinfo  {journal} {Rev. Mod. Phys.}\ }\textbf {\bibinfo {volume} {79}},\ \bibinfo {pages} {135} (\bibinfo {year} {2007})}\BibitemShut {NoStop}%
\bibitem [{\citenamefont {O’Brien}\ \emph {et~al.}(2009)\citenamefont {O’Brien}, \citenamefont {Furusawa},\ and\ \citenamefont {Vučković}}]{OBrien_2009}%
  \BibitemOpen
  \bibfield  {author} {\bibinfo {author} {\bibfnamefont {J.~L.}\ \bibnamefont {O’Brien}}, \bibinfo {author} {\bibfnamefont {A.}~\bibnamefont {Furusawa}},\ and\ \bibinfo {author} {\bibfnamefont {J.}~\bibnamefont {Vučković}},\ }\bibfield  {title} {\bibinfo {title} {Photonic quantum technologies},\ }\href {https://doi.org/10.1038/nphoton.2009.229} {\bibfield  {journal} {\bibinfo  {journal} {Nature Photonics}\ }\textbf {\bibinfo {volume} {3}},\ \bibinfo {pages} {687–695} (\bibinfo {year} {2009})}\BibitemShut {NoStop}%
\bibitem [{\citenamefont {Kimble}(2008)}]{Kimble_2008}%
  \BibitemOpen
  \bibfield  {author} {\bibinfo {author} {\bibfnamefont {H.~J.}\ \bibnamefont {Kimble}},\ }\bibfield  {title} {\bibinfo {title} {The quantum internet},\ }\href {https://doi.org/10.1038/nature07127} {\bibfield  {journal} {\bibinfo  {journal} {Nature}\ }\textbf {\bibinfo {volume} {453}},\ \bibinfo {pages} {1023–1030} (\bibinfo {year} {2008})}\BibitemShut {NoStop}%
\bibitem [{\citenamefont {Wehner}\ \emph {et~al.}(2018)\citenamefont {Wehner}, \citenamefont {Elkouss},\ and\ \citenamefont {Hanson}}]{Wehner_2018}%
  \BibitemOpen
  \bibfield  {author} {\bibinfo {author} {\bibfnamefont {S.}~\bibnamefont {Wehner}}, \bibinfo {author} {\bibfnamefont {D.}~\bibnamefont {Elkouss}},\ and\ \bibinfo {author} {\bibfnamefont {R.}~\bibnamefont {Hanson}},\ }\bibfield  {title} {\bibinfo {title} {Quantum internet: A vision for the road ahead},\ }\href {https://doi.org/10.1126/science.aam9288} {\bibfield  {journal} {\bibinfo  {journal} {Science}\ }\textbf {\bibinfo {volume} {362}},\ \bibinfo {pages} {eaam9288} (\bibinfo {year} {2018})},\ \Eprint {https://arxiv.org/abs/https://www.science.org/doi/pdf/10.1126/science.aam9288} {https://www.science.org/doi/pdf/10.1126/science.aam9288} \BibitemShut {NoStop}%
\bibitem [{\citenamefont {Xu}\ \emph {et~al.}(2020)\citenamefont {Xu}, \citenamefont {Ma}, \citenamefont {Zhang}, \citenamefont {Lo},\ and\ \citenamefont {Pan}}]{RevModPhys.92.025002}%
  \BibitemOpen
  \bibfield  {author} {\bibinfo {author} {\bibfnamefont {F.}~\bibnamefont {Xu}}, \bibinfo {author} {\bibfnamefont {X.}~\bibnamefont {Ma}}, \bibinfo {author} {\bibfnamefont {Q.}~\bibnamefont {Zhang}}, \bibinfo {author} {\bibfnamefont {H.-K.}\ \bibnamefont {Lo}},\ and\ \bibinfo {author} {\bibfnamefont {J.-W.}\ \bibnamefont {Pan}},\ }\bibfield  {title} {\bibinfo {title} {Secure quantum key distribution with realistic devices},\ }\href {https://doi.org/10.1103/RevModPhys.92.025002} {\bibfield  {journal} {\bibinfo  {journal} {Rev. Mod. Phys.}\ }\textbf {\bibinfo {volume} {92}},\ \bibinfo {pages} {025002} (\bibinfo {year} {2020})}\BibitemShut {NoStop}%
\bibitem [{\citenamefont {{Silvestri}}\ \emph {et~al.}(2024)\citenamefont {{Silvestri}}, \citenamefont {{Yu}}, \citenamefont {{Str{\"o}mberg}}, \citenamefont {{Hilweg}}, \citenamefont {{Peterson}},\ and\ \citenamefont {{Walther}}}]{Silvestri_2024}%
  \BibitemOpen
  \bibfield  {author} {\bibinfo {author} {\bibfnamefont {R.}~\bibnamefont {{Silvestri}}}, \bibinfo {author} {\bibfnamefont {H.}~\bibnamefont {{Yu}}}, \bibinfo {author} {\bibfnamefont {T.}~\bibnamefont {{Str{\"o}mberg}}}, \bibinfo {author} {\bibfnamefont {C.}~\bibnamefont {{Hilweg}}}, \bibinfo {author} {\bibfnamefont {R.~W.}\ \bibnamefont {{Peterson}}},\ and\ \bibinfo {author} {\bibfnamefont {P.}~\bibnamefont {{Walther}}},\ }\bibfield  {title} {\bibinfo {title} {{Experimental observation of Earth’s rotation with quantum entanglement}},\ }\href {https://doi.org/10.1126/sciadv.ado0215} {\bibfield  {journal} {\bibinfo  {journal} {Science Advances}\ }\textbf {\bibinfo {volume} {10}},\ \bibinfo {eid} {eado0215} (\bibinfo {year} {2024})}\BibitemShut {NoStop}%
\bibitem [{\citenamefont {{Hilweg}}\ \emph {et~al.}(2022)\citenamefont {{Hilweg}}, \citenamefont {{Shadmany}}, \citenamefont {{Walther}}, \citenamefont {{Mavalvala}},\ and\ \citenamefont {{Sudhir}}}]{Hilweg_2022}%
  \BibitemOpen
  \bibfield  {author} {\bibinfo {author} {\bibfnamefont {C.}~\bibnamefont {{Hilweg}}}, \bibinfo {author} {\bibfnamefont {D.}~\bibnamefont {{Shadmany}}}, \bibinfo {author} {\bibfnamefont {P.}~\bibnamefont {{Walther}}}, \bibinfo {author} {\bibfnamefont {N.}~\bibnamefont {{Mavalvala}}},\ and\ \bibinfo {author} {\bibfnamefont {V.}~\bibnamefont {{Sudhir}}},\ }\bibfield  {title} {\bibinfo {title} {{Limits and prospects for long-baseline optical fiber interferometry}},\ }\href {https://doi.org/10.1364/OPTICA.470430} {\bibfield  {journal} {\bibinfo  {journal} {Optica}\ }\textbf {\bibinfo {volume} {9}},\ \bibinfo {pages} {1238} (\bibinfo {year} {2022})}\BibitemShut {NoStop}%
\bibitem [{\citenamefont {Council}(2023)}]{ERC}%
  \BibitemOpen
  \bibfield  {author} {\bibinfo {author} {\bibfnamefont {E.~R.}\ \bibnamefont {Council}},\ }\href {https://cordis.europa.eu/project/id/101071779} {\bibinfo {title} {{GRAVITES – Gravitational interferometry with entangled states in optical fibers}}},\ \bibinfo {howpublished} {ERC Grant Agreement No. 101071779} (\bibinfo {year} {2023})\BibitemShut {NoStop}%
\bibitem [{\citenamefont {{Mieling}}\ and\ \citenamefont {{Hudelist}}(2025)}]{Mieling_2025}%
  \BibitemOpen
  \bibfield  {author} {\bibinfo {author} {\bibfnamefont {T.~B.}\ \bibnamefont {{Mieling}}}\ and\ \bibinfo {author} {\bibfnamefont {M.}~\bibnamefont {{Hudelist}}},\ }\bibfield  {title} {\bibinfo {title} {{Fiber optics in curved space-times}},\ }\href {https://doi.org/10.1103/PhysRevResearch.7.013162} {\bibfield  {journal} {\bibinfo  {journal} {Physical Review Research}\ }\textbf {\bibinfo {volume} {7}},\ \bibinfo {eid} {013162} (\bibinfo {year} {2025})}\BibitemShut {NoStop}%
\bibitem [{\citenamefont {{Neumann}}\ \emph {et~al.}(2022)\citenamefont {{Neumann}}, \citenamefont {{Selimovic}}, \citenamefont {{Bohmann}},\ and\ \citenamefont {{Ursin}}}]{Neumann_2022}%
  \BibitemOpen
  \bibfield  {author} {\bibinfo {author} {\bibfnamefont {S.~P.}\ \bibnamefont {{Neumann}}}, \bibinfo {author} {\bibfnamefont {M.}~\bibnamefont {{Selimovic}}}, \bibinfo {author} {\bibfnamefont {M.}~\bibnamefont {{Bohmann}}},\ and\ \bibinfo {author} {\bibfnamefont {R.}~\bibnamefont {{Ursin}}},\ }\bibfield  {title} {\bibinfo {title} {{Experimental entanglement generation for quantum key distribution beyond 1 Gbit/s}},\ }\href {https://doi.org/10.22331/q-2022-09-29-822} {\bibfield  {journal} {\bibinfo  {journal} {Quantum}\ }\textbf {\bibinfo {volume} {6}},\ \bibinfo {pages} {822} (\bibinfo {year} {2022})}\BibitemShut {NoStop}%
\bibitem [{\citenamefont {Ortega}\ \emph {et~al.}(2023)\citenamefont {Ortega}, \citenamefont {Fuenzalida}, \citenamefont {Selimovic}, \citenamefont {Dovzhik}, \citenamefont {Achatz}, \citenamefont {Wengerowsky}, \citenamefont {Shiozaki}, \citenamefont {Neumann}, \citenamefont {Bohmann},\ and\ \citenamefont {Ursin}}]{Ortega:23}%
  \BibitemOpen
  \bibfield  {author} {\bibinfo {author} {\bibfnamefont {E.~A.}\ \bibnamefont {Ortega}}, \bibinfo {author} {\bibfnamefont {J.}~\bibnamefont {Fuenzalida}}, \bibinfo {author} {\bibfnamefont {M.}~\bibnamefont {Selimovic}}, \bibinfo {author} {\bibfnamefont {K.}~\bibnamefont {Dovzhik}}, \bibinfo {author} {\bibfnamefont {L.}~\bibnamefont {Achatz}}, \bibinfo {author} {\bibfnamefont {S.}~\bibnamefont {Wengerowsky}}, \bibinfo {author} {\bibfnamefont {R.~F.}\ \bibnamefont {Shiozaki}}, \bibinfo {author} {\bibfnamefont {S.~P.}\ \bibnamefont {Neumann}}, \bibinfo {author} {\bibfnamefont {M.}~\bibnamefont {Bohmann}},\ and\ \bibinfo {author} {\bibfnamefont {R.}~\bibnamefont {Ursin}},\ }\bibfield  {title} {\bibinfo {title} {Spatial and spectral characterization of photon pairs at telecommunication wavelengths from type-0 spontaneous parametric downconversion},\ }\href {https://doi.org/10.1364/JOSAB.475583} {\bibfield  {journal} {\bibinfo  {journal} {J. Opt. Soc. Am. B}\ }\textbf {\bibinfo {volume} {40}},\ \bibinfo {pages}
  {165} (\bibinfo {year} {2023})}\BibitemShut {NoStop}%
\bibitem [{sm()}]{sm}%
  \BibitemOpen
  \href@noop {} {\bibinfo {title} {{See Supplemental Material at URL for the details of the photon source, interferometer characterization, procedures for photon measurements, and an explanation of the future differential measurement scheme.}}}\BibitemShut {Stop}%
\bibitem [{\citenamefont {{Mlejnek}}\ \emph {et~al.}(2017)\citenamefont {{Mlejnek}}, \citenamefont {{Kaliteevskiy}},\ and\ \citenamefont {{Nolan}}}]{mlejnek_2017}%
  \BibitemOpen
  \bibfield  {author} {\bibinfo {author} {\bibfnamefont {M.}~\bibnamefont {{Mlejnek}}}, \bibinfo {author} {\bibfnamefont {N.~A.}\ \bibnamefont {{Kaliteevskiy}}},\ and\ \bibinfo {author} {\bibfnamefont {D.~A.}\ \bibnamefont {{Nolan}}},\ }\href@noop {} {\bibinfo {title} {{Reducing spontaneous Raman scattering noise in high quantum bit rate QKD systems over optical fiber}}} (\bibinfo {year} {2017}),\ \Eprint {https://arxiv.org/abs/1712.05891} {arXiv:1712.05891 [quant-ph]} \BibitemShut {NoStop}%
\bibitem [{\citenamefont {{Kawahara}}\ \emph {et~al.}(2011)\citenamefont {{Kawahara}}, \citenamefont {{Medhipour}},\ and\ \citenamefont {{Inoue}}}]{KAWAHARA_2011}%
  \BibitemOpen
  \bibfield  {author} {\bibinfo {author} {\bibfnamefont {H.}~\bibnamefont {{Kawahara}}}, \bibinfo {author} {\bibfnamefont {A.}~\bibnamefont {{Medhipour}}},\ and\ \bibinfo {author} {\bibfnamefont {K.}~\bibnamefont {{Inoue}}},\ }\bibfield  {title} {\bibinfo {title} {{Effect of spontaneous Raman scattering on quantum channel wavelength-multiplexed with classical channel}},\ }\href {https://doi.org/10.1016/j.optcom.2010.09.051} {\bibfield  {journal} {\bibinfo  {journal} {Optics Communications}\ }\textbf {\bibinfo {volume} {284}},\ \bibinfo {pages} {691} (\bibinfo {year} {2011})}\BibitemShut {NoStop}%
\bibitem [{\citenamefont {{Mieling}}(2020)}]{Mieling_2020}%
  \BibitemOpen
  \bibfield  {author} {\bibinfo {author} {\bibfnamefont {T.~B.}\ \bibnamefont {{Mieling}}},\ }\bibfield  {title} {\bibinfo {title} {{On the influence of Earth's rotation on light propagation in waveguides}},\ }\href {https://doi.org/10.1088/1361-6382/ababb2} {\bibfield  {journal} {\bibinfo  {journal} {Classical and Quantum Gravity}\ }\textbf {\bibinfo {volume} {37}},\ \bibinfo {eid} {225001} (\bibinfo {year} {2020})}\BibitemShut {NoStop}%
\bibitem [{\citenamefont {{Mieling}}(2022)}]{Mieling_2022}%
  \BibitemOpen
  \bibfield  {author} {\bibinfo {author} {\bibfnamefont {T.~B.}\ \bibnamefont {{Mieling}}},\ }\bibfield  {title} {\bibinfo {title} {{Gupta--Bleuler quantization of optical fibers in weak gravitational fields}},\ }\href {https://doi.org/10.1103/PhysRevA.106.063511} {\bibfield  {journal} {\bibinfo  {journal} {\pra}\ }\textbf {\bibinfo {volume} {106}},\ \bibinfo {eid} {063511} (\bibinfo {year} {2022})}\BibitemShut {NoStop}%
\end{thebibliography}%
\end{document}


\preprint{APS/123-QED}

\title{Supplementary Materials}


\maketitle
\onecolumngrid

\section*{Photon source}

\begin{figure}[ht]
    \centering    \includegraphics[width=0.6\columnwidth]{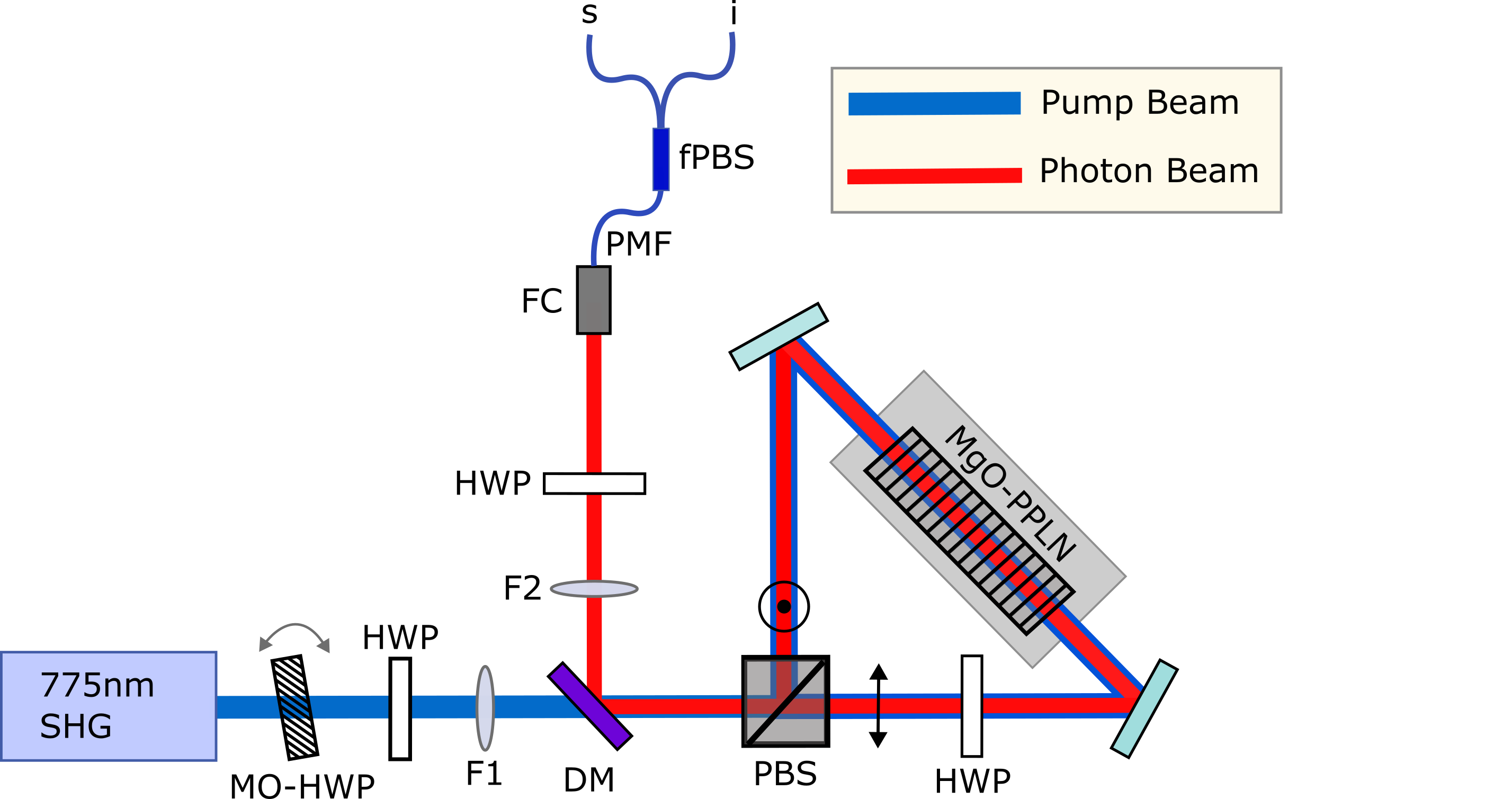}
    \caption{Schematic of the Type-0 photon source used in the experiment. A \SI{775.06}{\nano\meter} pump laser generated by second harmonic generation (SHG) from a \SI{1550.12}{\nano\meter} seed laser is sent through a multi-order HWP (MO-HWP), which is used to control the output optical state of the source. The polarization of the pump beam is set with another HWP before being focused by a \SI{200}{\milli\meter} aspheric lens (F1) towards the Sagnac interferometer. The crystal is housed in a temperature-controlled oven. Photons generated by the SPDC process are directed by the dichroic mirror (DM) and the photon beam is collimated with a \SI{250}{\milli\meter} aspheric lens (F2). A fixed HWP at 22.5 degrees inside the Sagnac loop rotates the input polarization along one arm to allow pumping of the crystal from both directions and generate the entangled state $\tfrac{1}{\sqrt 2}\left(\ket{HH} + \e^{\i\phi}\ket{VV}\right)$.  The photons are converted into the state $\tfrac{1}{\sqrt 2} \left(\ket{HV} + \ket{VH}\right)$ and coupled into polarization-maintaining fiber (PMF) via a fiber collimator (FC). A fiber PBS then separates the photons, directing the signal photons to the main interferometer and the idler photons to the nanowire detectors to serve as heralds for the signal.}
    \label{fig:type0_setup}
\end{figure}

To generate a high-brightness source of single photons for probing the interferometer, we built a Sagnac-geometry Type-0 spontaneous parametric down-conversion (SPDC) single-photon source, following the approach of \cite{Neumann_2022}.
We pump a 50-mm long MgO-doped periodically poled Lithium Niobate (PPLN) nonlinear crystal with a \SI{775}{\nano\meter} continuous-wave (CW) laser at \SI{1}{\watt}, generating photon pairs filtered to a \SI{100}{\giga\hertz} bandwidth at a rate of \SI{40}{\mega\hertz}.
The optical state out of the Sagnac loop is $\tfrac{1}{\sqrt 2}\left(\ket{HH} + \e^{\i\phi}\ket{VV}\right)$, where the relative phase $\phi$ between the polarizations is adjusted by rotating a multi-order half-wave plate (HWP) in the pump beam.
Another HWP and fiber polarizing beam splitter (fPBS) are used to convert the photon state into $\tfrac{1}{\sqrt 2}\left(\ket{HV} + \ket{VH}\right)$ and deterministically split the photon pair, with one of the photons sent into the 50-km interferometer and the herald photons to the superconducting nanowire single photon detector (SNSPD). The two photons are filtered using the same \SI{100}{\giga\hertz} dense wavelength-division multiplexers (DWDMs). A Hong-Ou-Mandel dip visibility of \SI{94.3}{\percent} is measured at the output of the source. 




\section*{Interferometer characterization with classical light}

To demonstrate the interferometer’s noise performance across the full frequency band, and to verify that the noise characteristics obtained using quantum light agree with those from classical light, we send CW laser light at \SI{1550}{\nano\meter} into the interferometer for characterization.
This light travels through the photon path and is directed to an independent homodyne detector after exiting the interferometer.

Supplementary Fig.~S2 shows the measured phase noise spectral density using photons and classical light, expressed in the unit of \unit{\radian\per\sqrt{\hertz}}. The red trace corresponds to the 160-hour single-photon interferometer measurement, matching the red trace in Fig.~2 of the main article.
The photon sensitivity achieves \SI{4.42e-3}{\radian\per\sqrt{\hertz}} at \SI{0.1}{\hertz}.
The blue trace shows the phase noise measured with \SI{1550}{\nano\meter} CW light when the interferometer is locked using the control laser field, achieving approximately \SI{6e-4}{\radian\per\sqrt{\hertz}} at \SI{0.1}{\hertz}. The black trace shows the dark noise of the homodyne detector.

\begin{figure*}[h]
\centering
\includegraphics[width=1.0\textwidth]{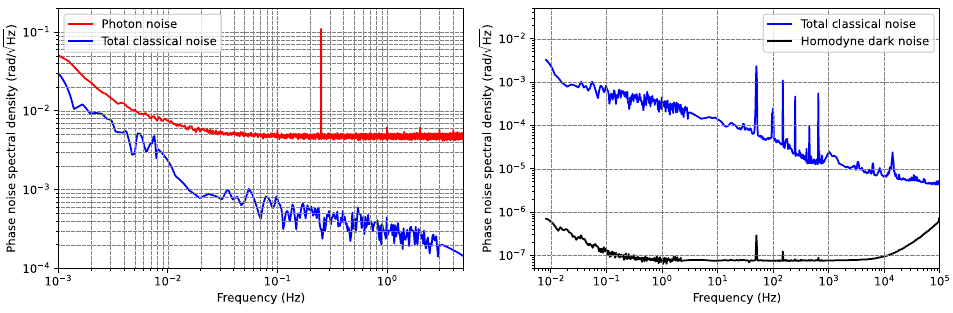}
\caption{Phase noise spectrum densities measured with quantum and classical light. 
(a) Comparison between the noise measured with single photons (red trace) and continuous classical light (blue trace). 
The quantum measurement is limited by photon shot noise at high frequencies. 
(b) Classical noise performance of the interferometer over the full frequency band from \SI{0.01}{\hertz} to \SI{100}{\kilo\hertz}, with customized homodyne detector dark noise as a reference (black trace).}
\label{fig:cw_spec}
\end{figure*}

\section*{Detailed procedures for photon measurements}
\label{sec:level3}

The main measurement presented in Fig.~2 of the main text consists of two continuous measurement runs with identical experimental setups, lasting \SI{91.2}{\hour} and \SI{68.8}{\hour}, respectively. Their phase noise spectra, shown separately in Supplementary Fig.~S3 highlight the apparatus' stability, consistency, and repeatability across different days.
\begin{figure*}[h]
    \centering
    \subfloat[\SI{91.2}{\hour} measurement period]{%
        \label{fig:fig1}%
        \includegraphics[width=0.49\columnwidth]{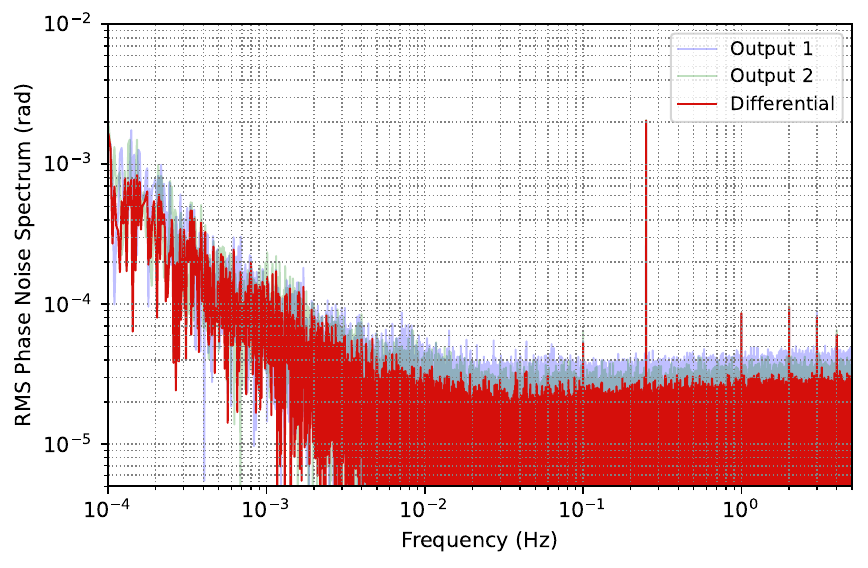}%
    }
    \hfill
    \subfloat[\SI{68.8}{\hour} measurement period]{%
        \label{fig:fig2}%
        \includegraphics[width=0.49\columnwidth]{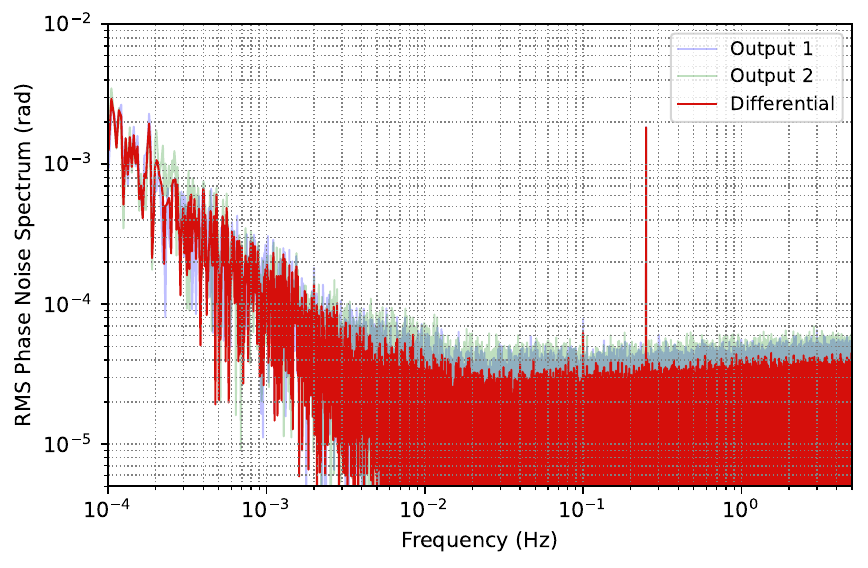}%
    }
    \caption{Phase noise spectrum of photon measurements lasting \SI{91.2}{\hour} and \SI{68.8}{\hour}, respectively, showing the repeatability across different measurements. The dim blue and green traces show the phase sensitivity calibrated from heralded single-photon counts at each interferometer output.
    The red trace shows the half-difference between the phase signals obtained from the two output ports of the interferometer. The peak at \SI{0.25}{\hertz} corresponds to the injected dither tone for calibration; The
    peak at \SI{0.1}{\hertz} represents the injected signal with an expected amplitude of \SI{6.48e-05}{\radian}, clearly resolved above the noise floor.}
    \label{fig:sidebyside}
\end{figure*}
The time-series data from the post-processing lock-in analysis are presented in Supplementary Fig.~S4. 
The measured interferometer phase is multiplied by a sine-wave reference signal, and the result is passed through a low-pass filter.
Shown are the low-pass filtered in-phase (I), quadrature (Q), and amplitude components of the demodulated signal corresponding to the injected dither at \SI{0.25}{\hertz}, measured over a period of \SI{91.2}{\hour}.
The phase of the reference signal is adjusted to eliminate the Q-component. 
The mean value of the I-component is recorded, with an expected value of \SI{2.10e-3}{\radian}.
%
\begin{figure*}[h]
    \centering\includegraphics[width=0.9\linewidth]{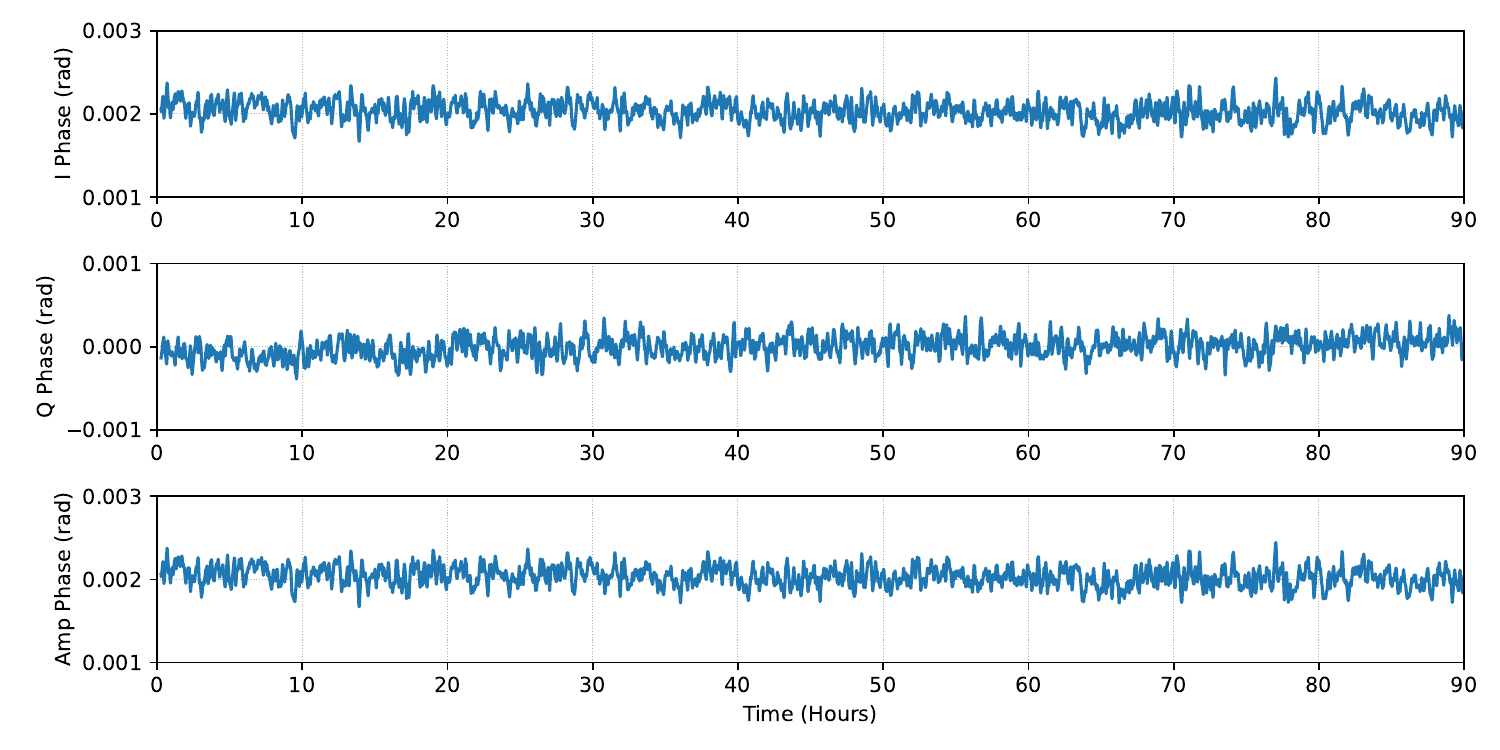}
    \caption{Lock-in analysis of the injected signal at \SI{0.25}{\hertz}.}
    \label{fig:fig4}
\end{figure*}

\section*{Explanation for future  differential measurement scheme}
\label{sec:level4}

The gravitational phase shift in Eq.~(1) of the main text is rigorously established for classical light. For entangled quantum fields, predictions from quantum field theory in curved spacetime (QFTCS) yield the same phase shift~\cite{Mieling_2023}; however, experimental validation of QFTCS remains elusive.

In a similar experimental setup with a vertical height difference between the two interferometer arms, the extracted quantity corresponding to $S_g$ in the main text would be the differential phase $\Delta\phi_{g, \mathrm{quantum}} - \Delta\phi_{g, \mathrm{classical}}$, where $\Delta\phi_{g, \mathrm{classical}} = \Delta\phi_{g}$ denotes the phase shift due to the gravitational redshift experienced by classical light, from the feedback control signal used to stabilize the interferometer; $\Delta\phi_{g, \mathrm{quantum}}$ is the gravity-induced phase shift experienced by the path-entangled quantum optical field within the framework of general relativity.
This difference reflects potential discrepancies between the gravitational coupling of classical and quantum states of light, providing a precision test of QFTCS that can either confirm its validity or reveal deviations indicative of new physics beyond the current theoretical consensus.
Practically, $\delta\phi_{g, \mathrm{classical}}$ can be determined by measurements with a CW field, combined with high-precision calibrations of the height difference and fiber length according to Eq.~(1). Because the gravitational redshift on a classical laser field is well understood, the evaluated $\delta\phi_{g, \mathrm{classical}}$ can, in principle, be used to recover $\delta\phi_{g, \mathrm{quantum}}$ in its entirety if desired.

\bibliography{references}